\documentclass[preprints,article,accept,moreauthors,pdftex]{Definitions/mdpi} 

\usepackage{amssymb}
\usepackage{amsmath}

\def\eps{\varepsilon}
\def\folgt{\quad\Rightarrow\quad}

\usepackage{color}
\usepackage{soul}
\usepackage{bbding}

\firstpage{1} 
\pubvolume{xx}
\issuenum{1}
\articlenumber{5}
\pubyear{2025}
\copyrightyear{2025}
\history{Received: October  19, 2026; Accepted: date; Published: date}

\Title{Gini coefficient of scientists and universities}

\Author{Klaus M. Frahm, $^{1}$\orcidA{} and 
Dima~L.~Shepelyansky$^{1,*}$\orcidB{},}

\AuthorNames{Klaus M. Frahm and Dima L. Shepelyansky}

\address{%
$^{1}$ \quad Univ Toulouse, CNRS, Laboratoire de Physique Th\'eorique,  
Toulouse, France\\

}

\corres{Correspondence: dima@irsamc.ups-tlse.fr; Tel. +33-56155-60-68 (D.L.S.)}

\abstract{Using the public database of Google Scholar
article citations of scientists and universities
we apply the algorithm for constructing Lorenz curves
to the resulting citation distributions.
Using Lorenz and Pareto curves, broadly used
for economic analyses of wealth inequality across countries, 
we show that this approach gives a useful
insight of citation distributions of world-leading scientists
characterized by the Gini coefficient $G$ obtained from the
Lorenz curves. For articles of 15 Nobel laureates 
(including A. Einstein and R. Feynman)
we find the average $G=0.81$ with 50\% of articles capturing only
2.7\% of all citations and the  top 10\% (1\%) of articles
capturing 71\% (30\%) of all citations.
Furthmore, we determine for world-leading universities
the Gini coefficient via the citations of affiliated scientists.
We show that about 65 percent of all citations
belonging to a university (e.g. Princeton, Cambridge)
are generated only by
10 percent of their university scientists
while 50\% of scientists produce only
3\% of all university citations.
The citation distributions are well described
by the concept of Rayleigh-Jeans thermalization
and condensation with similarities traced 
to the world wealth inequality.
}
  

\begin{document}

\section{Introduction}
\label{sec1}

The scientific analysis of article citations of scientists
is an active research field with a variety of
methods of statistical investigation
(see, e.g., \cite{bornmann2008} and references therein).
Some indicators have become rather popular,
such as the h-index \cite{hirsch},
although some of their limitations are actively discussed
(see, e.g., \cite{bornmann2007} and references therein).
New methodological approaches are still under development
\cite{bornmann2026,hungary2}.
Here we propose a new characterization of
a scientist's articles and of universities
based on their citations. The citations are obtained
from the open public database of Google Scholar
between 11 and 17 September 2026 \cite{scholar}.

Our approach
is based on the algorithm for constructing the Lorenz curve \cite{lorenz}
with related Gini coefficient \cite{gini}.
We also construct the Pareto distribution of citations \cite{pareto}.
These characteristics are broadly used in economic studies of wealth inequality
across countries worldwide (see, e.g., 
\cite{piketty1,piketty2,boston,yakovenko2}). 
In  this contex, the distribution of wealth
of households is described by the Lorenz curve 
\cite{lorenz,boston} which presents the dependence of cumulated 
normalized wealth fraction $0\leq w \leq 1$ on the cumulative normalized 
fraction of the
population or households $0 \leq h \leq 1$.
The equipartition of wealth is given by
the diagonal $w=h$ and the doubled area between the diagonal
and the Lorenz curve $w(h)$ determines the Gini coefficient
$0 \leq G \leq 1$  \cite{gini,boston}.
A detailed description of the construction
of the Lorenz curve and Gini coefficient is for example available in
\cite{lcurve,ginicoef}. 
A remarkable example is the Lorenz curve of the whole world in 2021 
which is characterized by a high Gini coefficient $G=0.843$ \cite{piketty2},
indicating a huge wealth inequality in the world. Indeed,
according to \cite{piketty2} in 2021 
50\% of the world population owns only 2\% of total wealth while 
10\% (1\%) of the population owns 75\% (38\%) of total wealth.

The application of the Lorenz curve approach to the article citations
is straightforward: for the articles of a scientist, the number of
citations $c_m$ per article $m$ is analogous to the wealth $w_m$ of an 
individual household $m$, so that we assume the equivalence  $c_m = w_m$.
The individual household weight of each article is chosen as $h_m= 1/N_a$
where $N_a$ is the total number of articles of a scientist.
We consider that $c_m$ is the total number 
of article citations at the time of data collection.
The normalization is performed using the total number of citations $N_c$
of a scientist for all of his/her articles.
For universities, we collect from \cite{scholar}
all scientists linked to a given university
and consider the university citations as
the total number of citations $c_m$ of a given scientist
(being assimilated to an effective contribution $w_m=c_m$
of a scientist to the ``citation wealth'' of his/her university).
The normalization is performed using the total number of citations
$N_c$ of all $N_s$ scientists linked to a given university.
The effective quantity analogous
to the weight $h_m$ of an individual household (scientist) $m$ is 
chosen as $h_m =1/N_s$ using the total number of scientists $N_s$ of a 
given university. 
As for the case of wealth,
the Lorenz curve of citations is obtained as the
normalized cumulative citation fraction
$c=w$ as a function of the normalized cumulative
number $h$ of articles (for a given scientist)
or a cumulated number $h$ of scientists (for a  given university).
In the following, we use symbols $w, h$ for the Lorenz and Pareto curves 
describing citations of scientists' articles and universities.

To the best of our knowledge,
there is only rather limited use of the Lorenz curve description of citations
in the scientometric literature (see, e.g., \cite{hungary1}
where the Lorenz curves were obtained for
article distributions over countries
for six scientometric journals with Gini coefficients $G \approx 0.7$).

It is important to note that the Lorenz curve
and Gini coefficient describe quantities normalized
by the total number of citations. In this sense, they
characterize the structure and importance of
articles of a scientist or
contribution of scientists within the citation
structure of a given university.
However, the characteristics 
obtained and discussed here 
are not directly related to the total
number of citations of articles of a scientist
or citations of scientists of a given university.

The article is organized as follows: data sets are described in 
Section \ref{sec2};
properties of Rayleigh-Jeans thermalization and condensation
are presented in Section~\ref{sec3};
construction rules of  Lorenz and Pareto curves are given in
Section~\ref{sec4}; results are described in Section~\ref{sec5};
the basic assumptions underlying the validity of thermalization  are
highlighted in Section~\ref{sec6}
and the discussion of the results is given in Section~\ref{sec7}.
The Appendix presents specific additional figures and 
analyzes the thermalization properties for the specific case 
with only one energy integral of motion. 
Other additional figures are given in the Supplementary Material.

\section{Data sets} 
\label{sec2}

To perform the Lorenz curve analysis of the citations of articles of scientists
we select 25 scientists with names given in Table~\ref{tab1}. Among
these scientists are 12 Nobel Prize laureates in Physics:
A. Einstein (1921), R. Feynman (1965), K. von Klitzing (1985), W. Ketterle (2001),
K. Novoselov (2010), S.Haroche (2012), D.Wineland (2012), D.Thouless (2016),
G.Parisi (2021), A.Zeilinger (2022), J.Hopfield (2024), M. Devoret (2025)
(ordered by the prize year). The sample also includes R. Myerson
(Nobel Memorial Prize in Economic Sciences in 2007),
Jennifer Doudna (Nobel Prize laureate in Chemistry in 2020),
D. Baker (Nobel Prize laureate in Chemistry in 2024),

The 25 selected scientists also include leading scientists
in network science (A.-L.Barabasi, M.Newman, Reka Albert),
in nonlinear dynamics and chaos (C.Grebogi, A.Pikovsky),
in quantum computing (P. Zoller) and machine learning (Y. LeCun).
These 25 names also include our laboratory colleague C. Sire
and the two authors of this article.

For each scientist, the citation counts of his/her articles
are collected from Google Scholar \cite{scholar}
including only articles with at least one citation. 
The number of articles $N_a$ is different for each scientist
ranging from 83 to 1300 (Baker) with, for example, the following 
numbers of articles: 498 (Einstein), Feynman (126), 622 (Barabasi), 
658 (LeCun), 695 (Doudna), 882 (Novoselov). 
The total number of citations $N_c$
for each of the 25 scientists is given in Table~\ref{tab1}.

\begin{table}[H]
\caption{Citation (wealth) 
inequality indicators for 25 scientists ($*$ marks Nobel laureates). 
The columns correspond to name, 
Gini coefficient $G$, the total number of citations $N_c$, 
share of total citations (wealth) held by the bottom 50\%,
the top 10\%, and the top 1\% of articles (households) for each scientist. 
}
\label{tab1}
\begin{center}
\begin{tabular}{lcrccc}
\toprule
\textbf{Name} & \textbf{Gini coefficient} $G$ & $N_c$ &
\textbf{Bottom 50\%} & \textbf{Top 10\%} & \textbf{Top 1\%} \\
\midrule
Albert R. & 0.926 &  141958 & 1.1\% & 91.8\% & 53.5\% \\
Baker D. * & 0.676 &  223834 & 8.2\% & 54.3\% & 15.9\% \\
Barabasi A.-L. & 0.881 &  335483 & 1.0\% & 81.7\% & 38.1\% \\
Devoret M. * & 0.738 &  87418 & 3.8\% & 58.1\% & 18.7\% \\
Doudna J. * & 0.817 &  171105 & 2.4\% & 72.0\% & 34.8\% \\
Einstein A. * & 0.883 &  200742 & 1.5\% & 84.0\% & 39.9\% \\
Feynman R. * & 0.850 &  140985 & 0.9\% & 74.5\% & 31.4\% \\
Frahm K.~M. & 0.594 &  2757 & 9.8\% & 40.5\% & 6.8\% \\
Grebogi C. & 0.759 &  51612 & 5.3\% & 65.0\% & 30.7\% \\
Haroche S. * & 0.777 &  51499 & 2.4\% & 64.2\% & 18.8\% \\
Hopfield J. * & 0.868 &  98732 & 1.9\% & 80.7\% & 45.8\% \\
Ketterle W. * & 0.753 &  80966 & 3.3\% & 60.8\% & 20.4\% \\
von Klitzing K. * & 0.793 &  39889 & 4.5\% & 70.0\% & 42.1\% \\
LeCun Y. & 0.913 &  495413 & 0.9\% & 87.4\% & 54.0\% \\
Myerson R.~B. * & 0.876 &  52467 & 1.2\% & 81.7\% & 34.1\% \\
Newman M. & 0.878 &  279347 & 1.0\% & 83.3\% & 27.6\% \\
Novoselov K. * & 0.926 &  458485 & 0.5\% & 89.4\% & 59.1\% \\
Parisi G. * & 0.799 &  116359 & 4.1\% & 70.8\% & 35.4\% \\
Pikovsky A. & 0.836 &  45593 & 3.3\% & 76.7\% & 48.5\% \\
Shepelyansky D. & 0.619 &  15199 & 10.9\% & 49.5\% & 10.8\% \\
Sire C. & 0.599 &  6028 & 10.0\% & 43.3\% & 6.0\% \\
Thouless D. * & 0.830 &  43012 & 2.5\% & 75.2\% & 29.5\% \\
Wineland D. * & 0.783 &  91834 & 1.8\% & 65.3\% & 13.1\% \\
Zeilinger A. * & 0.810 &  128825 & 2.0\% & 69.4\% & 22.6\% \\
Zoller P. & 0.732 &  152921 & 5.2\% & 60.2\% & 22.7\% \\
\bottomrule
\end{tabular}
\end{center}
\end{table}

\begin{table}[H]
\caption{Citation (wealth) inequality indicators for eight universities. 
The columns correspond to name, 
Gini coefficient $G$, the total number of (reduced) citations $N_c$ of
university scientists, number of scientists $N_s$ (with at least 
500 citations),
share of total citations (wealth) held by the bottom 50\%,
the top 10\%, and the top 1\% of affiliated scientists (household)
for each university. Only affiliated scientists with at least 500 citations 
are taken into account and for them the number of citations has been 
reduced by 500 such that minimal values are (close to) zero. 
For comparison the bottom row provides the characteristics of world wealth 
inequality in 2021 
taken from \cite{piketty2,jsp}. (The $N_c$ column shows the World GDP in 
billions of USD and the $N_s$ column gives the number of world countries.)
}
\label{tab2}
\begin{center}
\begin{tabular}{lcrrccc}
\toprule
\textbf{Name} & \textbf{Gini coefficient} $G$ & $N_c$ & $N_s$ &\textbf{Bottom 50\%}
& \textbf{Top 10\%} & \textbf{Top 1\%} \\
\midrule
Bonn & 0.752 &  9130679 & 1265 & 4.5\% & 62.3\% & 22.6\% \\
Cambridge & 0.775 &  34292866 & 4234 & 3.6\% & 65.5\% & 19.7\% \\
ENS & 0.686 &  2382410 & 285 & 7.1\% & 55.0\% & 13.7\% \\
ESPCI & 0.680 &  1249979 & 236 & 7.3\% & 53.9\% & 16.9\% \\
ETH & 0.775 &  18390366 & 3176 & 3.6\% & 65.4\% & 18.8\% \\
Princeton & 0.802 &  25200088 & 2454  & 2.6\% & 68.8\% & 20.9\% \\
Sapienza & 0.752 &  15764575 & 2397 & 5.8\% & 65.3\% & 30.0\% \\
Technion & 0.681 &  6997732 & 1273 & 7.0\% & 52.5\% & 18.8\% \\
\midrule
World & 0.843 & 101560 & 180 & 2\% & 75\% & 38\% \\
\bottomrule
\end{tabular}
\end{center}
\end{table}

We also selected 8 leading universities
in different countries presented in Table~\ref{tab2}. 
For each university we determine the number of scientists
affiliated with this university
who have at least $N_{\min}=500$ citations and 
we choose a reduced citation number $w_m=c_m-N_{\min}$ for them 
when constructing Lorenz and Pareto curves using the wealth values $w_m$. 
In this way, the used minimal wealth values (reduced citation numbers) 
are always (close to) zero and such a shift can be absorbed in a corresponding 
shift of the RJ model spectra (by choosing $E_0=0$ there; see next section). 
Furthermore, the values of 
$N_c$ in Table~\ref{tab2} corresponds to the sum of all reduced citations 
numbers ($N_c=\sum_m w_m=\sum_m (c_m-N_{\min})$). 
The reason for selecting this minimal citation threshold $N_{\min}$ 
is that universities have many graduate students and post-docs
who have relatively few citations. For the example of Princeton University 
this procedure selects $N_s=2454$ scientists (with at least 500 citations) 
out of $5973$ scientists having at least 1 citation. 
At the same time the articles published by
the affiliated students may in fact have been produced
in other universities before they arrived at Princeton.
Also, we are more interested in a scientific production
produced at the university by regular researchers and not by the production
of low-cited articles of students. 
We note that such an increase
of the citation threshold
from 1 to 500, and assuming reduced citation wealth values, 
does not change significantly 
the Lorenz and Pareto curves (e.g.
for Princeton $G$ is reduced from
$G=0.885$ at $N_{\min}=1$ to $G=0.802$ at $N_{\min}=500$,
see discussion below in Section~\ref{sec3}). 

The number of scientists $N_s$ (with $c_m\ge N_{\min}$) 
for each university is also provided in Table~\ref{tab2}.
The number of articles $N_a$ of certain scientists are given in the text above.
All Figures with Lorenz and Pareto curves provide these values of $N_a$ or 
$N_s$ (noted as $N$ in the figure panels).

\section{Rayleigh-Jeans thermalization and condensation}
\label{sec3}

In this section, we present the concept
of Rayleigh-Jeans (RJ) thermalization and condensation
which we apply to describe the citation distributions 
of scientists and universities and in particular 
their Lorenz and Pareto curves.

In the recent work \cite{jsp}
the Wealth Thermalization Hypothesis (WTH)
was proposed with to explain 
the striking wealth inequality observed for countries
and the whole world \cite{piketty1,piketty2}.
The WTH assumes that various types of links and nonlinear 
interactions among members of society (agents) in a given country
lead to a dynamical thermalization over layers of wealth in a society
with a corresponding social stratification  \cite{stratif}.
In the  WTH approach the wealth values $w_m$ of households $h_m$
are associated with
energies $E_m$  of states $m$ (or related frequencies).
The thermalization sets in from the classical dynamical
nonlinear equations leading to chaos. This process is assumed to   
have two conserved quantities being total system
energy $E = \sum_m E_m \rho_m$ and probability norm 
$\eta = \sum_m \rho_m =1$ (rescaled number of agents). 
Here $\rho_m$ are the time-averaged squared amplitudes and occupation 
probabilities of each oscillator mode $m$ \cite{jsp}.  
According to the well-known results
of thermodynamics of classical fields \cite{landau,zakharovbook}
the steady-state distribution of probabilities (or fractions)
for the RJ thermalization is:
\begin{equation}
\rho_m = \frac{T}{E_m-\mu} \; ({\rm RJ}) .
\label{eqrj}
\end{equation}
Here $\rho_m$ represent averaged stationary 
probabilities at certain wealth states 
 $0 \leq m < N$ with wealth/energies $w_m=E_m$. The 
two parameters correspond to the system temperature $T$ and 
the chemical potential $\mu(T)$ which are 
determined from the values of the total energy $E$ and the norm $\eta=1$
from the two equations above  (see, e.g., Refs. 
\cite{jsp,rmtprl,ourfiber,complex} and below for more details).
Within this analogy between citation number of an article $c_m$ and
wealth value $w_m$ we assume the equivalence $c_m=w_m=E_m$ 
(or $w_m=c_m-N_{\min}$ for the case of University citations where 
$N_{\min}=500$). 
In the following, we mainly use $w_m$ to denote citation values $c_m$ 
(or reduced citation values $c_m-N_{\min}$).

A remarkable feature of the thermal RJ distribution (\ref{eqrj})
is the appearance of RJ condensation at low values of the total energy $E$: 
a large fraction of total probability is condensed at the low-energy 
states $E_m$.
The condensation fraction can be very close to unity $\eta=1$
(see e.g., \cite{jsp,ourfiber,complex}) and it explains naturally 
the emergence of 
poor and oligarchic phases in the wealth inequality \cite{jsp,complex}.
The existence of RJ condensation was first pointed out by 
H.~Fr\"ohlich in 1968 in the context of electrical modes in 
biological cells \cite{frohlich1,frohlich2}
(however, in these works there is a logical gap as pointed out 
in \cite{jsp,ourfiber,complex}).
For multimode optical fibers the RJ condensation
was found in numerical simulations \cite{picozzi1} and 
experimentally observed in
\cite{wabnitz,picozzi2}. In fact, the RJ condensation is part of
a more general phenomenon of statistical mechanics known as the
constraint-driven condensation \cite{trizac,satya,marsili}.
This phenomenon describes
such diverse physical systems as
coalescence in granular media, jamming in traffic, 
gelation in networks \cite{satya}
and financial data analysis \cite{marsili}.

In Refs. \cite{jsp,complex}, it was shown that the WTH approach
based on the RJ thermalization and condensation describes well 
the distributions for the wealth inequality in countries; 
gross domestic product (GDP) distribution over world countries;
market capitalization of companies on the stock exchanges of Hong Kong,
Shanghai, London; Bitcoin transactions and world trade flows.
Applications of RJ thermalization and condensation (\ref{eqrj})
to other systems can be also found in \cite{election} and references therein.
Here we assume the hypothesis that there is also RJ thermalization and 
condensation 
for citations of scientists and universities being analogous to those
of the WTH \cite{jsp,stratif,complex}. Indeed, it is known that
links between scientists form a complex network
(see, e.g., \cite{newman2001,newman2006,newmanbook})
and nonlinear interactions there lead to the RJ thermalization and condensation
\cite{jsp,stratif}). Such interactions appear via scientific publications,
conferences, joint discussions, and recently via Internet-based exchanges.

\begin{figure}[h]
\begin{center}
\includegraphics[width=0.95\columnwidth]{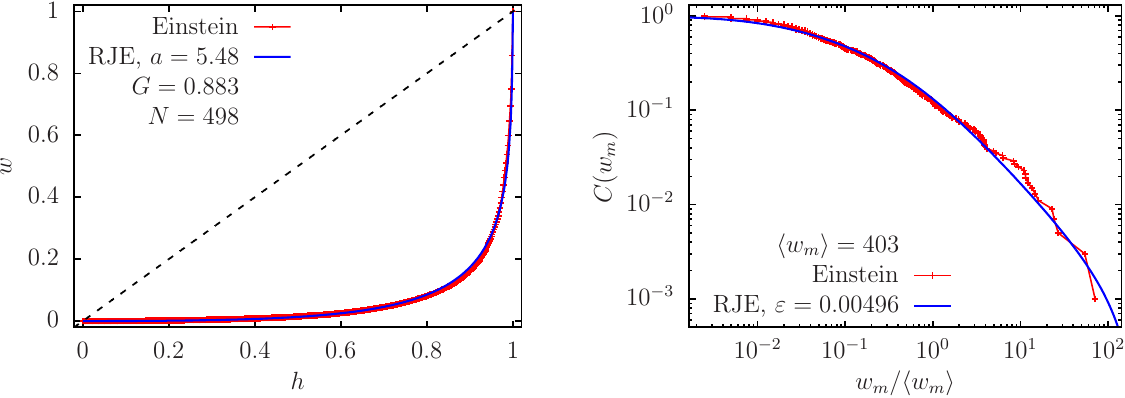}%
\end{center}
\caption{\label{fig1}
\label{fig_einstein1_multi}
{\em Left:} Comparison of the citation Lorenz curve for 
Einstein (red data curve) with 
the theoretical RJE model (blue curve). 
The $x$-axis corresponds to the 
cumulative fraction of households/papers ($h$) and the $y$-axis to
the cumulative fraction of wealth/citations ($w$). 
The dashed black line 
corresponds to the line of perfect equipartition $w=h$. 
The corresponding values of $a$ and $\eps$ have been determined by 
minimizing the geometrical curve distance between both curves and 
such that they have the same Gini coefficient $G=0.883$. 
{\em Right:} Pareto curve $C(w_m)$ for the same data (red 
data curve) where 
$C(w_m)$ represents the fraction of articles having more citations 
than $w_m$ (analogous to wealth). 
The $x$-axis corresponds to rescaled 
values $w_m/\langle w_m\rangle$ where 
$\langle w_m\rangle$ is the average citation number per article. 
The blue curve corresponds to the 
theoretical RJE Pareto curve with the same parameters as in the left 
panel (all parameters are given in the figure). 
Note that for the theoretical RJE model, we have $w_m\in[0,1]$ 
and $\langle w_m\rangle=\eps$ (see text and \cite{complex} for details).
Here and in all other figures, $N$ denotes the number of articles $N_a$
of a scientist (as here) or the number of scientists $N_s$ at a given 
university. 
}
\end{figure}

\section{Construction of Lorenz and Pareto curves}
\label{sec4}

For real data, we consider $N$ wealth/citation values $w_m$ 
ordered as $w_0\le w_1\le\ldots\le w_{N-1}$ (here 
$N=N_a$, $w_m=c_m$ for scientists or $N=N_s$, $w_m=c_m-N_{\min}$ 
for the university cases with $N_{\min}=500$). 
Then the associated Lorenz curve $(h(m),w(m))$ 
for $N+1$ data points $m=0,1,\ldots,N$ is simply constructed 
as $h(m)=m/N$ and $w(m)=\sum_{k=0}^{m-1} w_m/(N w_s)$ 
where $w_s=\sum_{k=0}^{N-1} w_k/N$ is the average wealth/citation value. 
Here $h(m)$ represents the cumulative fraction of households (papers of 
a scientist or scientists affiliated to a university) at index $m$ and 
$w(m)$ is the cumulative normalized citation (wealth) fraction such 
that $h(0)=w(0)=0$ and $h(N)=w(N)=1$. In this construction, we assume 
that each paper/scientist of a scientist/university has the same uniform 
household fraction $1/N$.

For the RJ theory, we consider an ordered model spectrum $E_k$ of $N$ levels 
with $0=E_0<E_1<\ldots< E_{N-1}$ for large values of $N$ (e.g. $N=10^4$, 
see also below). Here, each level corresponds to a given agent $k$ 
with associated wealth $E_k$. We attribute to each agent a thermalized 
RJ probability $\rho_k=T/(E_k-\mu)$ as in (\ref{eqrj}) for a given 
energy value $E\in]E_0,E_{N-1}[$ which fixes the parameters $T$ and 
$\mu$ by the two implicit equations: $1=\sum_k \rho_k$ and 
$E=\sum_k E_k\rho_k$. Then, we define the RJ Lorenz curve as 
$N+1$ points $(h(m),w(m)$ (for $m=0,1,\ldots,N$)
with $h(m) =\sum^{m-1}_{k=0} \rho_k$ 
and $w(m) = \sum^{m-1}_{k=0} E_k \rho_k/E$, again satisfying 
$h(0)=w(0)=1$ and $h(N)=w(N)=1$. 
This curve is invariant with respect to a simple energy 
rescaling $E_m\to\alpha E_m$, $E\to \alpha E$ (and similarly for $\mu$ 
and $T$). Thus, the relevant 
parameter is the rescaled energy $\varepsilon = E/B$ 
where $B=E_{N-1}$ is the energy bandwidth of the spectrum.
In this theoretical RJ approach, it is assumed that each university 
scientist (or article of a scientist)
is represented by a certain number
of thermalized agents distributed over the 
energy levels $E_k$ being in the energy band $[0,B]$. 

Concerning, the choice of the RJ energy spectrum, 
the simplest assumption is that the density of energy states
$\nu(k) =d k/d E_k$ is constant so that
$0 \leq E_k=k/N < 1$ for $k=0,1,2,\ldots,(N-1)$. This model,
called the RJ standard (RJS) model, was analyzed in detail 
in \cite{ourfiber,jsp}. 
For sufficiently small values of $\eps$ it has been shown that 
RJ condensation takes place where 
a significant fraction of probability is concentrated on a 
small number of modes at minimal $k$ values. This description
provides Lorenz curves 
that agree on a qualitative level with  Lorenz curves from real data. 

However, to obtain a more precise description, we use a more detailed 
model called the RJ extended (RJE) model with 
$E_k =(e^{ak/N}-1)/(e^a-1)$ with $a$ being 
an additional parameter besides $\eps$. The limit $a\to 0$ gives the RJS model 
with $E_k=k/N$ and for $a>0$ the density of states $\nu(E_m)$ is decreasing at 
high energies as $\nu(E_k)=dk/dE_k = N(e^a-1)/[a(1+(e^a-1)E_k]$. The global 
energy scale here is chosen such that $B=E_{N-1}\approx 1$ (such that 
$E\approx \eps$). For the wealth inequality, the decay of $\nu$ at 
high energies is rather natural since
there are fewer rich people  at high wealth/energy values.
We consider that such a situation is also plausible for the 
density distribution of citations 
where highly cited articles are rarer. 
In this work, we  present results (mostly) for the RJE model
since the RJS model gives only a qualitative agreement
with real data.

To recover the Lorenz curve from the RJ theory (\ref{eqrj}),
one can use the above procedure for large but finite values of $N$, 
e.g. $N=10000$. Here, we use the continuous limit $N\to\infty$ for 
which it is possible to work out closed analytic expressions 
\cite{complex} for 
the Lorenzcurve $w(h)$ (depending on $\eps$, $\mu$ which is fixed by a 
certain implicit equation) and also for other quanties ($G$ and Pareto curve; 
see below). These analytic RJ expressions agree very well the finite $N$ 
RJ curves provided $N$ is sufficiently large (identical on graphical 
precision for $N\ge 10^4$). 
We remind, that in the RJE approach with (very) large values of $N$, 
multiple agents (many values of the spectral index $k$) correspond 
to a given individual data point for real citation values $w_m$ (one 
fixed value of $m$). 

To compare the RJE model with real Lorenz curves, the parameter 
$\eps$ is fixed by the condition that the RJ and real Lorenz curve have the 
same Gini coefficient $G$. In addition, for the RJE model the geometrical 
curve distance between both curves is minimized to determine the optimal 
choice for the parameter $a$ (under the constraint of 
respecting the first condition). 
This procedure is the same as the one used in \cite{jsp,complex}. 

\begin{figure}[h]
\begin{center}
\includegraphics[width=0.95\columnwidth]{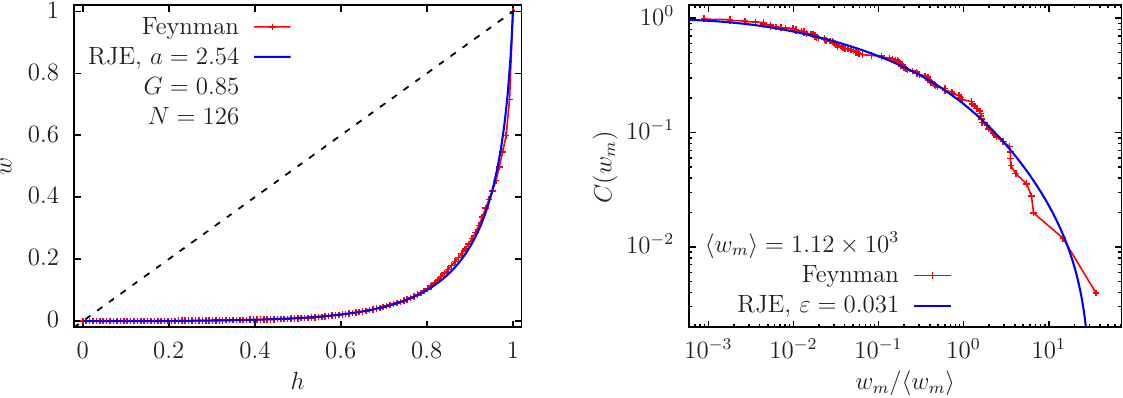}%
\end{center}
\caption{\label{fig2}
\label{fig_feynman1_multi}
Same as Fig.~\ref{fig1} for R. Feynman.
}
\end{figure}

\begin{figure}[h]
\begin{center}
\includegraphics[width=0.95\columnwidth]{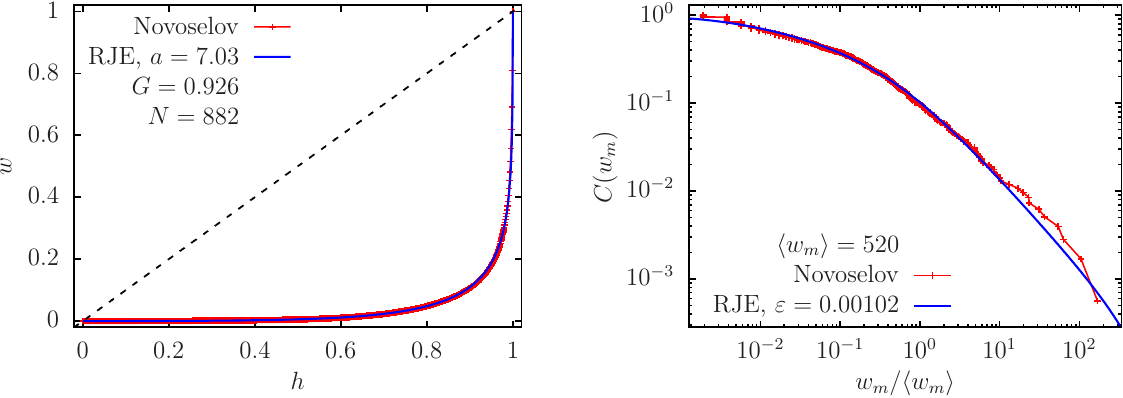}%
\end{center}
\caption{\label{fig3}
\label{fig_novoselov1_multi}
Same as Fig.~\ref{fig1} for K. Novoselov.
}
\end{figure}

In addition to the Lorenz curve it is also useful to introduce  
another quantity that emphasizes the details for the very 
rich population/article fraction. 
In economics, this is the cumulative distribution function (CDF) 
$C(w_m)$, or Pareto distribution, that gives the fraction of households 
(cited articles or scientis, companies, people, 
countries, etc.) having a wealth (citation number) greater than 
$w_m$. This quantity is 
typically used in economics (see, e.g., \cite{pareto,yakovenko2}) 
and it can be directly obtained 
by drawing $C(w_m)=1-h(m)$ versus $w_m$ (for both cases 
of real data and RJ models). Usually  the Pareto distribution
is described by an algebraic decay law. 
In our work, we simply use the notation 
{\em Pareto distribution (or curve)}  for the CDF $C(w_m)$, 
for more general cases even if there is no 
simple power law. 

In contrast to the Lorenz curve, the Pareto curve 
is not scale independent and in order to compare real data with 
possible RJE curves, it is necessary to fix the global energy/biomass 
scale either by the maximal or the average value of $w_m$. Here, we show 
$C(w_m)$ as a function of $w_m/\langle w_m\rangle$ where for the 
RJS/RJE models, we simply have $\langle w_m\rangle=\eps$ (since $B=1$ and 
$E_m=w_m$). For real data the average value $\langle w_m\rangle$ 
is more reliable since the 
maximal value of $w_m$ for a given data set may be subject to 
strong statistical fluctuations (note that 
$\max(w_m)=1$ for the RJS/RJE models 
since $B=1$).

\begin{figure}[h]
\begin{center}
\includegraphics[width=0.95\columnwidth]{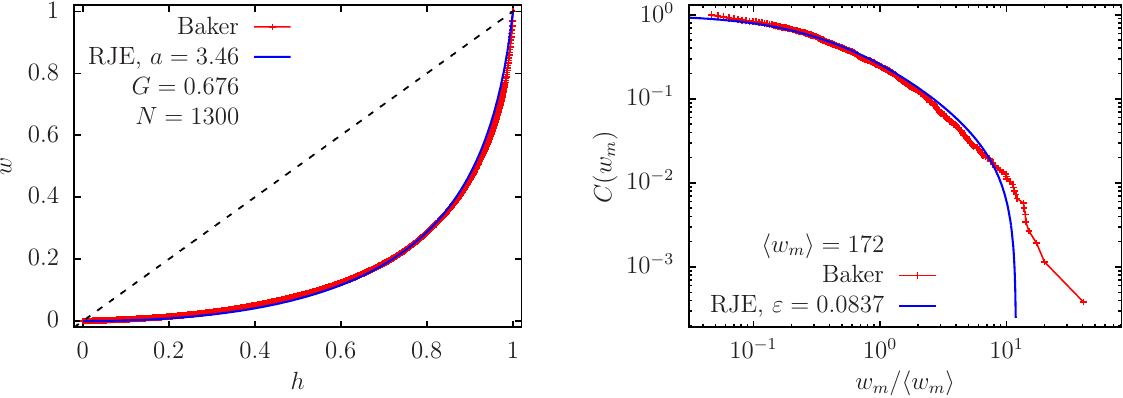}%
\end{center}
\caption{\label{fig4}
\label{fig_baker1_multi}
Same as Fig.~\ref{fig1} for D. Baker.
}
\end{figure}

\begin{figure}[h]
\begin{center}
\includegraphics[width=0.95\columnwidth]{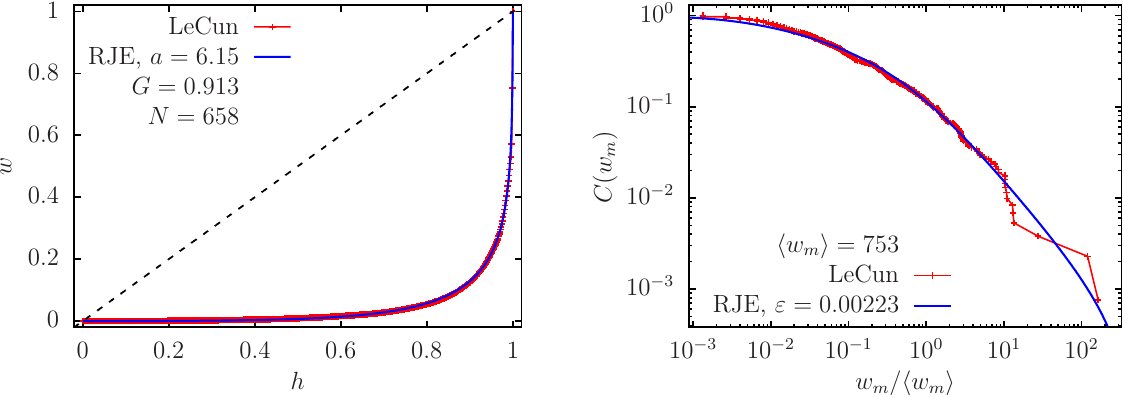}%
\end{center}
\caption{\label{fig5}
\label{fig_lecun1_multi}
Same as Fig.~\ref{fig1} for Y. LeCun.
}
\end{figure}

\section{Results} 
\label{sec5}

\subsection{Citations of articles of  25 scientists}
\label{subsec5.1}

The global characteristics of Lorenz curves and Gini coefficients
for citations of articles of 25 selected scientists are
presented in Table~\ref{tab1}. The three highest Gini values
are for R. Albert, K. Novoselov, Y. LeCun indicating
that a large fraction of a scientist's total citations
is produced by a small fraction of their articles.
Indeed, for these 3 scientists we have respectively only 1.1\%, 0.5\%, 0.9\%
of total citations for the bottom 50\% of articles,
91.8\%, 89.3\%, 87.4\% of citations for the top 10\% of  articles,
and 53.5\%, 59.1\%, 54\% for the top 1\% of articles.
This citation inequality of articles is even more striking
compared to the world wealth inequality described in \cite{piketty2}
(see percent numbers given in the text above in Section~\ref{sec1}).

For illustration, we have included in Table~\ref{tab1} 
some scientists with modest citations values $N_c$ and 
they have lower values 0.594, 0.599, 0.619 of the Gini coefficient 
(KMF, C. Sire, DS) and for them the citation inequality is less pronounced.
However, overall there is no direct correlation between
total number of citations $N_c$ and Gini coefficient.

For the 15 Nobel laureates the average 
characteristics of Table~\ref{tab1} are:
$G= 0.812 \pm 0.0617$;
the citation fraction of the bottom 50\% of articles is $2.73\% \pm 1.84\%$ 
and the top 10\% (1\%) citation fraction 
is $71.4\% \pm 9.58\%$ ($30.8\% \pm 12.3\%$).
These average numbers are quite comparable to the corresponding 
values for the world wealth inequality in 2021 \cite{piketty2,jsp}. 
Furthmore, for the 15 Nobel laureates, the average number of citations is
$N_c = 132410 \pm 102676$ and the average number of articles
is $N_a = 506 \pm 313$ corresponding to an average 
of $261.6$ citations per article.
At the same time the maximal numbers of citations
per article are 51800 (R. Albert), 87652 (K. Novoselov),
122643 (Y. LeCun) and 28657 for A. Einstein.

For all 25 scientists, the corresponding average Gini coefficient is 
$G = 0.797 \pm 0.0945$; the citation fraction for the bottom 50\% of articles 
is $ 3.58\% \pm 3.02\%$ 
and the top 10\% (1\%) citation fraction is 
$ 70\% \pm 13.7\%$ ($30.4\% \pm 14.7\%$). 
Here, we have $N_c = 140499 \pm 128405$ and 
$N_a =  458 \pm 283$. These numbers are similar to those
for of Nobel laureates.
We attribute this to the fact that the selected set
of scientists includes the leading experts in their scientific
fields (KMF, CS, DS are added to represent
more typical cases of researchers).

In Fig.~\ref{fig1} we show the Lorenz and Pareto curves
of citations of $N_a=498$ articles of A. Einstein.
The RJE Lorenz curve with optimal parameters $\eps=0.0049$
and $a=5.48$ is very close to the real data Lorenz curve. 
The very small value of $\eps$ shows that
there is a condensate of articles with low citations
(only 1.5\% of citations for the bottom 50\% of all articles).
Top 10\% (1\%) of articles capture 84.0\% (39.9\%)
of all citations (see Table~\ref{tab1}).
The most cited article is about
Einstein-Podolsky-Rosen (EPR) pairs \cite{epr}
with 28657 citations. The citation number 
of this article is boosted
by the development of quantum information and communication.

It is possible to consider the case when the minimal number
of citations of articles of A. Einstein
is restricted to $M=10$ (with a subsequent shift $w_m=c_m-10$ 
for the computation of Lorenz and Pareto curves) 
instead of $M=1$ as shown in Fig.~\ref{fig1}. 
We can assume that in this case some 
not very important notes/publications
are excluded. Our results show that this procedure
only slightly reduces $G$ from $0.883$ at $M=1$ to
$0.87$ at $M=10$ without significant modification
of curve shapes. Therefore, in all cases of Table~\ref{tab1}, 
we present results only for $M=1$. 

We note that Google Scholar also provides articles
citing a given article for a given year. For example, we 
consider the $N_a =355$ articles published in 2001 citing 
the famous EPR paper \cite{epr} and we construct the Lorenz and 
Pareto curves for them using as wealth quantity their own 
citation number $w_m=c_m$ (these articles, citing \cite{epr} in 2001, 
accumulate in total $N_c=45029$ citations). 
The resulting Lorenz and Pareto curves are shown in Appendix Fig.~\ref{figA1} 
and they are well described by the RJE model 
with $\eps = 0.0111$, $a=4.45$ corresponding to the Gini coefficient $G=0.861$.
The small value of $\eps$ results in a significant RJ condensation
with the bottom 50\% of articles capturing
only 1.7\% of total citations of \cite{epr} in 2001,
the top  10\% (1\%) of  articles capturing 79\% and 40\%
of all citations. This example shows that the RJE model 
also decribes the citation statistics of papers citing 
a given (famous) article in a given year.

The Lorenz and Pareto curves for citations of articles of
R. Feynman are shown in Fig.~\ref{fig2}.
In this case the RJE Lorenz and Pareto curves also agree well 
with the real data curves. There is  a bit higher deviation
from at the tail of Pareto curve that we
attribute to low statistics at the smallest values of $C(w_m)$
due to a relatively modest number of articles of Feynman ($N_a=126$).

The Lorenz and Pareto curves for citations of
articles of K. Novoselov are presented in Fig.~\ref{fig3}
agreeing well with the RJE model. Here 
the number of articles $N_a=882$ is significantly higher. 
The effective system energy is very low with $\eps = 0.00102$
such that the condensate phase is very large
corresponding to the highest Gini value $G=0.926$
(see other parameters in Table~\ref{tab1}).

The case of citations of articles of D. Baker
is significantly different and also the number of articles $N_a = 1300$ 
is extremely. Here, the Gini coefficient $G=0.676$
is the lowest among the Nobel laureates. 
As it is shown in Fig.~\ref{fig4}, the RJE model still gives a good description
of the Lorenz curve data and for the Pareto data for the range 
$C(w_m) > 0.01$ (and with some deviations at the tail for smaller values 
of $C(w_m)$). A possible reason of the somewhat
low value of $G$ can be related to very broad applications of protein
structures in various fields of chemical and biological sciences.

For articles of Y. LeCun the RJE Lorenz and Pareto curves
also demonstrate a good agreement with the real data.
Here the three most-cited articles already capture about 50\% of all citations
corresponding to a very high condensation of articles with low citations
and concentration of articles with high citations
(analogous to the oligarchic phase in wealth inequality \cite{jsp,complex}).
Here, the Gini coefficient has a very  high value $G=0.913$. 
We note that for Novoselov and LeCun, two cases with large $G$, also 
the optimal values of the RJE parameter $a\approx 6$-$7$ is rather high 
confirming that the most strongly cited articles are very rare. 

In addition to the five scientists discussed above and in 
Figs.~\ref{fig1}-\ref{fig5}, 
we have in Table~\ref{tab1} 11 additional Nobel laureates having rather high
Gini coefficients in the range $G = 0.738$ (M. Devoret) to $G= 0.876$ 
(R. Myerson); these scientists also have a rather large bottom-50\% 
fraction of articles with a low fraction of total citations 
in the range from 0.5\% (K. Novoselov) to 4.5\% (K. von Klitzing).
At the same time for this group of scientists
the top 10\% of articles capture from 58\% to 81.7\% of total citations.

Another 9 leading scientists represent such different areas of science as:
network science (R.Albert, A.-L.Barabasi, M. Newman);
nonlinear dynamics, chaos,  synchronization (C. Grebogi, A. Pikovsky);
quantum computing (P. Zoller).
Also KMF, C. Sire, DS represent more typical cases of researchers in physics.

The Lorenz and Pareto curves for the other 20 scientists of Table~\ref{tab1}
are given in Supplementary Material file supmat.pdf;
the corresponding figures in alphabetic order are: 
Fig.~S1 (Albert), Fig.~S2 (Barabasi),
Fig.~S3 (Devoret), Fig.~S4 (Doudna), Fig.~S5 (Frahm),
Fig.~S6 (Grebogi), Fig.~S7 (Haroche), Fig.~S8 (Hopfield),
Fig.~S9 (Ketterle), Fig.~S10 (Klitzing),
Fig.~S11 (Myerson), Fig.~S12 (Newman),
Fig.~S13 (Parisi), Fig.~S14 (Pikovsky), Fig.~S15 (Shepelyansky),
Fig.~S16 (Sire), Fig.~S17 (Thouless), Fig.~S18 (Wineland),
Fig.~S19 (Zeilinger), Fig.~S20 (Zoller).
The RJE curves in Fig.~S1 to Fig.20 agree globaly very well 
with the real data of Lorenz and Pareto curves. 
In Fig.~S9 for articles of W. Ketterle, we also show as illustration 
Lorenz and Pareto curves for the RJS model (green curve) corresponding 
to the limit $a\to 0$ which only gives a qualitative description of real 
data with some significant deviations in comparison 
to the RJE model (blue curve). We note that here the optimal RJE 
parameter $a=2.38$ is quite small such that the difference between RJE 
and RJS curves are quite modest (however clearly visible).
For other cases, with larger optimal values of $a\approx 4$-$7$ the deviations 
between RJS and RJE (and therefore between RJS and real data) 
are even stronger. For this reason, we do not include RJS curves in the other 
figures. 

\begin{figure}[h]
\begin{center}
\includegraphics[width=0.95\columnwidth]{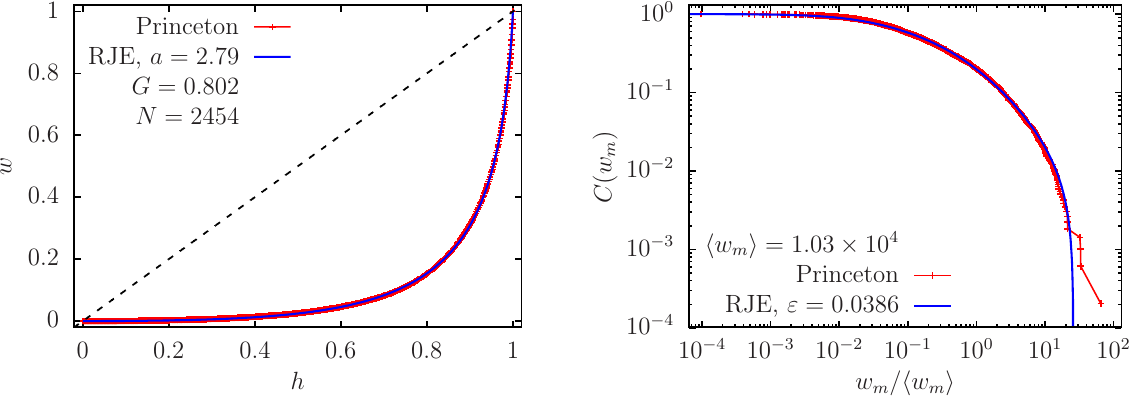}%
\end{center}
\caption{\label{fig6}
\label{fig_PrincetonM500_multi}
Same as Fig.~\ref{fig1} for Princeton University.
}
\end{figure}

\subsection{Citations of eight universities}
\label{subsec5.2}

Here, we present the statistical analysis of citations
of scientists affiliated with a specific university.
At Google Scholar certain universities (e.g. Princeton University)
are associated with scientists that put this university
as their affiliation on the web page \cite{scholar}
and each such scientist is shown with the total number of his/her citations.
Therefore, it is possible to compute citation Lorenz curves for universities 
where the affiliated scientists correspond to the household variable 
(with identical uniform weight for each scientist at a given university) 
and their total numbers of (reduced) citations 
correspond to the wealth variable. 
To be more precise, as already explained above, we only consider university 
affiliated scientists with at least $N_{\min}=500$ citations 
($c_m\ge N_{\min}$) and use for them 
a shifted wealth variable (reduced citation number) $w_m=c_m-N_{\min}$ 
to compute Lorenz and Pareto curves such that minimal wealth 
values $w_m$ are (close) to zero. We use this minimal threshold due 
the presence of many young researchers (usually PhD
students or post-docs) who have very low citation numbers (down to $c_m=1$).
Many such students or post-docs come to a given university
from outside and bring with them citations of their publications
that quite often are associated with other universities.
In Appendix Fig.~\ref{figA2}, we show for the example of 
Princeton University, that the threshold of $N_{\min}=500$ 
does not significantly affect the Lorenz curve. In particular, 
there is only a 10\% decrease of the Gini coefficient
from $G=0.883$ (no threshold) to $G=0.802$ at $N_{\min}=500$.

In this work, we consider the cases of 8 universities 
listed in Table~\ref{tab2}. 
This table presents the same quantities as Table~\ref{tab1} which 
are the Gini coefficient $G$, the total number of (reduced) citations 
$N_c$ obtained by the $N_s$ considered scientist, the value of $N_s$ itself, 
and the citation fractions of bottom 50\%, top 10\%, top 1\% scientists. 
As illustration and for comparison, Table~\ref{tab2} also contains 
an additional row with 
these quantities for the world wealth data of 2021 \cite{piketty2,jsp}.

In Table~\ref{tab2}, we use short names of universities,
the associated full names are: University of Bonn (Bonn),
University of Cambridge (Cambridge),
\'Ecole normale sup\'erieure de Paris (ENS),
\'Ecole sup\'erieure de physique et de chimie industrielles 
de la ville de Paris (ESPCI),
Eidgen\"ossische Technische Hochschule Z\"urich (ETH),
Princeton University (Princeton),
Sapienza University of Rome (Sapienza),
Technion -- Israel Institute of Technology (Technion).

\begin{figure}[h]
\begin{center}
\includegraphics[width=0.95\columnwidth]{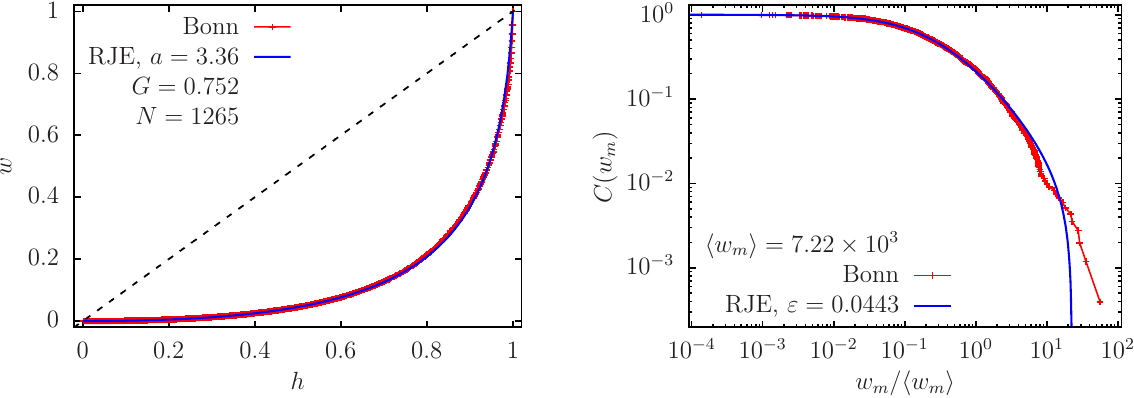}%
\end{center}
\caption{\label{fig7}
\label{fig_BonnM500_multi}
Same as Fig.~\ref{fig1} for Bonn.
}
\end{figure}

\begin{figure}[h]
\begin{center}
\includegraphics[width=0.95\columnwidth]{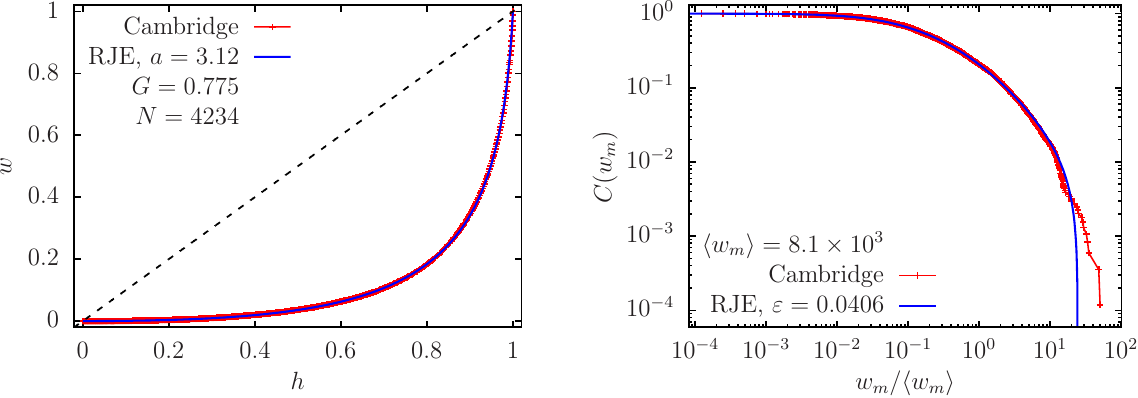}%
\end{center}
\caption{\label{fig8}
\label{fig_CambridgeM500_multi}
Same as Fig.~\ref{fig1} for Cambridge.
}
\end{figure}

The characteristics of Table~\ref{tab2} show that
Princeton University has the highest Gini coefficient 
$G=0.802$. The most striking feature is that 50\% of scientists
produce only 2.6\% of all university citations 
while 10\% (1\%) of all scientists produce
68.8\% (20.9\%) of all university citations.
The most cited scientist is Daniel Kahneman
with 666949 citations (prospect theory).
Albert Einstein is at 9th position with 201059 citations.
We note that the citations are collected for
the whole time of the functioning
of the university and over all scientists affiliated with it 
that are registered at \cite{scholar} and have at least 500 citations.

The Lorenz and Pareto curves for Princeton University
are shown in Fig.~\ref{fig6}. The RJE theory curves describe 
well the real data with only a small
deviation at the tail of the Pareto curve
where the statistical number of events is very low. 
Note that in Appendix Fig.~\ref{figA2}, the Princeton Lorenz curve 
is also shown (for the two cases $N_{\min}=1$ and $N_{\min}=500$) in 
a logarithmic scale on the $y$-axis and the agreement between the RJE and 
real data curve remains good for the full range of the cumulated wealth 
covering 7 orders of magnitude. 

\begin{figure}[h]
\begin{center}
\includegraphics[width=0.95\columnwidth]{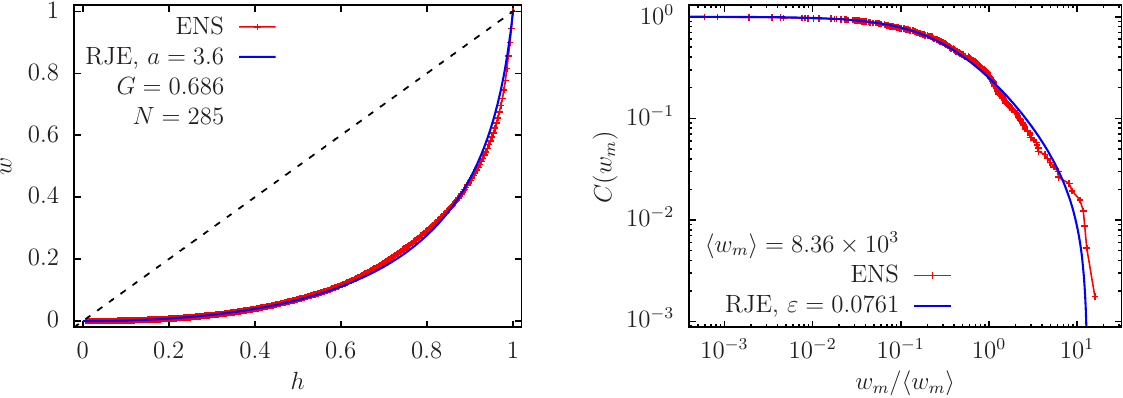}%
\end{center}
\caption{\label{fig9}
\label{fig_ENSM500_multi}
Same as Fig.~\ref{fig1} for ENS.
}
\end{figure}

\begin{figure}[h]
\begin{center}
\includegraphics[width=0.95\columnwidth]{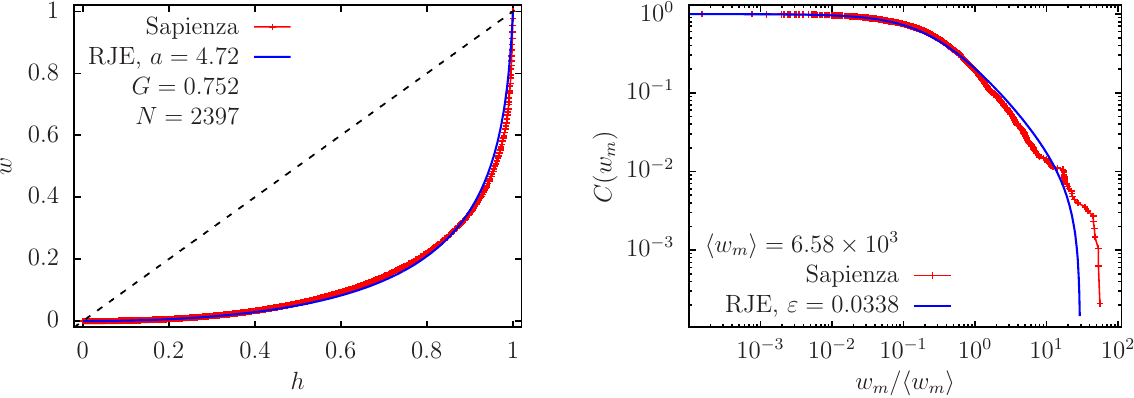}%
\end{center}
\caption{\label{fig10}
\label{fig_SapienzaM500_multi}
Same as Fig.~\ref{fig1} for Sapienza.
}
\end{figure}

Fig.~\ref{fig7} shows the Lorenz and Pareto curves
for the case of Bonn with a good agreement between
the RJE model and real data. Here the Gini coefficient
$G=0.752$ is slightly smaller as compared to Princeton.
The condensate fraction of 50\% of scientists with low citations
captures only 4.5\% of all university citations 
while 10\% of scientists captures 62.3\% of all citations.

The case of Cambridge, shown in Fig.~\ref{fig7}, is rather similar to 
Princeton with a good agreement between RJE and real data curves. 
The Gini coefficient
and other characteristics in Table~\ref{tab2}
are also similar to those of Princeton. As
for Bonn there are certain deviations between RJE theory and real data
at the tail of the Pareto curve
which we attribute to statistical fluctuations
due to a small number of events in this tail.

Fig.~\ref{fig9} also demonstrates a good agreement
between RJE theory and real data for Lorenz and Pareto curves for ENS. 
We note that the number of scientists for ENS is almost 15 times smaller 
as compared to Cambridge (see Table~\ref{tab2}).
The bottom condensate fraction of 50\% scientists captures here
a relatively elevated citation fraction of 7.1\%
and the fraction of the top 10\% scientists captures 55\% of all citations.
The case of ESPCI (see also SuppMat Fig.~S21) has also a rather small 
number of affiliated scientists 
and characteristics similar to those of ENS as it is shown in Table~\ref{tab2}.

For Sapienza the Lorenz and Pareto curves and their comparison
with the RJE theory are shown in Fig.~\ref{fig10};
there are certain deviations between theory
and data at the Pareto curve tail
attributed to a small number of events.
The characteristics of Sapienza in Table~\ref{tab2} are
relatively close to those of Bonn.
At the same time, the top 1\% of scientists captures
30\% of all citations which is the highest value for the 8 universities
presented in Table~\ref{tab2}.

In the Supplementary Material file supmat.pdf, 
we show the Lorenz and Pareto curves 
for ESPCI (Fig.~S21), ETH (Fig.~S22),
Technion  (Fig.~S23). In all these cases, 
the RJE and real data curves agree well (with usual modest deviations 
at the tail of the Pareto curves).

\begin{figure}[h]
\begin{center}
\includegraphics[width=0.95\columnwidth]{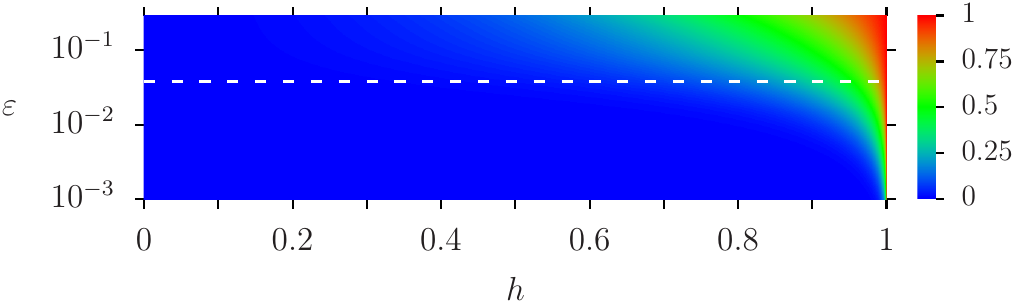}%
\end{center}
\caption{\label{fig11}
\label{fig_LorCol}
Color plot of cumulated wealth $w$ from the Lorenz curves of the RJE model 
for the parameter $a=2.79$. The $x$-axis corresponds to 
the fraction of households $h\in[0, 1]$ 
and the $y$-axis to the rescaled 
energy $\varepsilon=E/B\in[10^{-3}, \eps_C]$ in logarithmic representation 
where $\eps_C=0.293$ is the critical energy value for the transition 
from positive to negative temperature for $a=2.79$. 
The values of the colorbar represent the cumulated wealth fraction 
$w\in[0,1]$ at given values of $h$ and $\eps$. 
The white dashed line corresponds to the value 
$\eps=0.0386$ used in Fig.~\ref{fig_PrincetonM500_multi} 
for the blue RJE curve (with same value of $a$) being 
the best fit for the data of Princeton. 
Note that the color values along the dashed line correspond to the blue RJE 
curve in the left panel of Fig.~\ref{fig_PrincetonM500_multi}. 
}
\end{figure}

According to Fig.~\ref{fig6}, we know that for Princeton 
(case with threshold $N_{\min}=500$) the optimal RJE parameters are 
$\eps =0.0386$ and $a=2.79$ at which the RJE Lorenz curve agrees 
very well with the real data Lorenz curve of Princeton. 
It is interesting to analyze how the RJE Lorenz curve $w(h)$ evolves when 
the rescaled energy $\eps$ varies at a fixed value of $a$. 
In Fig.~\ref{fig11}, we show a color plot of $w(h,\eps)$ 
at $a=2.79$  
for $\eps\in[10^{-3},\eps_C]$ with a logarithmic scale for $\eps$ on the 
$y$-axis while $h\in[0,1]$ corresponds to the $x$-axis in normal scale. 
Here $\eps_C=0.293$ is the critical energy for the transition from 
positive to negative temperature. 
This Figure clearly illustrates that 
with a decrease of $\eps$ below the optimal Princeton value $\eps=0.0386$
the condensate phase is significantly increased while
above $\eps=0.0386$ the condensate phase is reduced but it still 
remains visible. 

It is useful to compute the average and standard deviation of 
the quantities of Table~\ref{tab2} with respect to 
the 8 universities which gives $G=0.738 \pm 0.0455$. 
The average citation fraction of bottom 50\% is $5.19\% \pm 1.73\%$ and 
the average citation fraction of top 10\% (1\%s) is 
$61.1\% \pm 5.91\%$ ($20.2\% \pm 4.47\%$). 
Furthermore, we have 
$N_c = 14176087 \pm 10779940$ for the average total university citation 
number and 
$N_s =1915 \pm 1313$ for the average university scientist number. 
Thus there are on average 7403 citations per scientist.
The highest number of citations per scientist is
10268 for Princeton. 

In our opinion, the most striking result of this subsection
is that globally the bottom 50\% fraction of university scientists
produces on average only 5.19\% of all university citations,
while the top 10\% (1\%) of scientists
produces 61\% (20\%) of all university citations.
For Princeton and Cambridge this citation inequality is 
even sharper with approximate values of 
3\% for the bottom 50\% of scientists
and 67\% (20\%) for the top 10\% (1\%) of university scientists.
We note that these numbers are rather similar
to those of the  wealth inequality in the world
(see \cite{piketty2,jsp}, introduction text above and 
also last extra row in Table~\ref{tab2}).

\section{Main assumptions for the validity of RJ thermalization and condensation}
\label{sec6}

In this work, we presented numerous examples of citation Lorenz curves 
for scientists and Universities where the optimal RJE fits agree well 
or even very well with the real data Lorenz curves. 
A critical reader might argue that this agreement is not very astonishing 
in view of the two free parameters $a$ and $\eps$ of the RJE model 
and that it is not sufficient to clearly establish that the citation 
distributions are indeed based on 
RJ thermalization with the distribution (\ref{eqrj}). 
However, the agreement also extends to the cumulated distribution 
function, which we call Pareto curves for simplicity. Here the tails, 
shown in a double logarithmic scale over several orders of magnitude, 
correspond to a very small region 
of the Lorenz curve (with $h\to 1$) such that the fit procedure does 
not force a strong agreement in this part. Furthermore, 
for the Pareto curves, we need to reintroduce the global energy scale 
(see discussion at the end of Section \ref{sec4}) given by the 
average $\langle w_m\rangle$ which is not an additional fit parameter 
but it is directly computed for both cases of real data and of the RJE model 
(where $\langle w_m\rangle=\eps$). Here the strong agreement for 
most Pareto curves over many orders of magnitude is indeed a strong 
argument in favor of the RJ approach.

We also 
mention, that other two parameter fit functions (e.g. for the Pareto curve) 
based on a power law and some hard or smooth transition from a non-power law 
(e.g. constant or exponential) to the power law regime (with the second 
parameter for the transition scale) do not work well for the real data 
citation Pareto curves shown in this work. In particular, for many cases 
of the real data Pareto curves, 
we do not see a straight line over a significant interval (in 
double logarithmic representation). We can mention that for the case 
of Bitcoin data analyzed in \cite{complex} there is a significant interval 
with a power law but also for this case the RJE fits, with rather high 
values of $a\approx 10-12$, are superior and actually valid for the 
complete data interval (also at the boundaries with significant deviations 
between power law and real data). 
Actually, in \cite{complex} it was shown that 
for large $a$ values the RJE model 
contains a power law for some limited interval. 

It is true that
we give here only heuristic arguments
for possible nonlinear couplings and interactions
between articles of a scientist
or interactions between university scientists
that can lead to chaos and thermalization.
However, here we can point out that for thermalization
of a gas of atoms in a 3D box almost any weak or
moderate interactions between atoms lead to their
thermal distribution \cite{landau}. 

Apart from this, we present the following arguments in
support of the RJ WTH theory:

A1) in \cite{jsp,complex} it was shown
that the obtained Lorenz curves
for wealth, GDP, London stock exchange
remain very stable on a scale of 30-50 years;
for given names (see arXiv:2608.06048)
such a stability is present on a scale of a century 
which is an important feature for a thermal distribution.

A2) RJ distributions describe the Lorenz curves of many 
different systems such as wealth inequality 
of households in various countries; market capitalization of 
companies at stock exchanges of Hong Kong, Shanghai, London;
GDP distribution over world countries  during 50 years;
world trade; Bitcoin transactions (see, e.g., \cite{jsp,complex});
given names statistics (arXiv:2608.06048);
distribution of energy consumption over countries (arXiv:2607.07315);
distribution of votes between parties of EU elections (arXiv:2607.15119);
distribution of wild land mammal biomass (arXiv:2610.09744);
distribution of citations described in this work. 
This demonstrates the universality of RJE Lorenz curves 
for many different systems which is also 
an important feature of a thermalization theory. 

A3) In all systems of point A2 above there is a strong condensation
in the shape of Lorenz curves when only a small fraction
of total wealth (or other characteristics) of about a couple of percent  is
captured by about 50 percent of total population,
while a small (oligarchic-like) fraction of population
captures about 50-70 percent of total wealth.
Such a condensation phenomenon naturally appears in the framework of
RJ thermalization when the total system energy is small
compared to the total energy dispersion (or band) of the system.

We think that these arguments provide 
a strong support for the RJ WTH concept.
This concept assumes an analogy between certain energy levels
corresponding to levels of social stratification
in a society \cite{stratif}. Also it is assumed that the system
evolution has two integrals of motion being total system
energy and total probability norm.
In fact in \cite{stratif}, for the case of system with turbulent flow, 
an indication appeared showing that the norm (number of particles)
conservation is not a strict condition. 
Indeed, in the Appendix, we analyze the case when the system has only one 
energy integral 
and the norm or number of particles is not conserved.
In this case, the RJ thermal distribution
is still given by the equation (\ref{eqrj})
but with zero chemical potential $\mu=0$. 
As shown in the Appendix, 
the Lorenz curves can still be
described by the RJ theory at $\mu=0$
but in such a case it is required
that the system energy spectrum
$E_m$ has a spectral gap with a minimal positive energy 
$\Delta_0=E_0>0$ (for the two integral case $E_0$ can have any 
sign but we need $\mu<E_0$ for $T>0$). 
A strong RJ condensation is
possible only at a big ratio $B/\Delta_0 \gg 1$ (with $B=E_{N-1}$ being 
the band width) 
and in such a case the real data Lorenz curves can also be well described by 
the RJ distribution at $\mu=0$ (see Appendix
Figs.~\ref{figA3}-\ref{figA6}).

Finally, we point out that not every 
system with its Lorenz curve can be described by the RJS or RJE thermal
distribution (\ref{eqrj}) with two integrals of motion.
As an example we refer to the statistical properties of
poker tournaments for which it was shown that the real data are nicely 
described by a statistical theory developed in \cite{poker1}.
The Lorenz curve of this statistical distribution
has a low Gini coefficient $G=0.365$ \cite{poker2} which is close to
the limiting coefficient $G=1/3$ for very high temperatures in the RJS 
model with essentially uniform probabilities $\rho_m$ \cite{jsp}.
The Lorenz and Pareto curves
from real data \cite{poker1} and from RJS and RJE theory
are shown in Appendix Fig.~\ref{figA7}. Their comparison shows that
the Lorenz curves from RJS and RJE models are practically the same 
(due to a small value of the optimal parameter $a$). 
Both theory curves are still quite close to the real 
Poker data \cite{poker1,poker2} Lorenz curve 
but there are visbible differences. At the same time, 
the associated Pareto curves, shown in Appendix Fig.~\ref{figA7} (right),
are drastically different between
the real data of \cite{poker1,poker2} and the corresponding 
the RJS and RJE Lorenz curves (with best possible fit to determine $a$ 
and matching Gini coefficient to fix $\eps$).
In fact the Pareto curve was given in the inset of Fig.3
of \cite{poker1} and the real Pareto data is well described by
the statistical theory developed in \cite{poker1}.
In a certain sense, this theory \cite{poker1}
links the statistical analysis of poker tournaments
with the statistical persistence problem at relatively weak (or absent)
interactions between players. This leads to another statistical description
not related to RJ thermalization even if the Gini coefficient 
$G=0.365$ is close to the equipartition limit of the RJS model 
(but this distribution is valid only for those
players who survived in a long tournament). 

\section{Discussion}
\label{sec7}

In this work we presented a statistical analysis of citations of  articles of a scientist
and citation production of universities obtained via
citations of affiliated scientists.
The analysis is performed
on the basis of the algorithm of Lorenz curve construction
broadly used in economics.
We find that the citations of leading scientists are characterized
by a high Gini coefficient
$G \approx 0.8$
and high the bottom 50\% fraction of articles capturing only a few percent of all citations
while a small top fraction of 10\% captures about 70\% of all citations.
For citations of universities we show
that bottom fraction of 50\% of scientists generates only about 3\%
of all citations while the top fraction 10\% of scientists
produce about 65\% of all university citations.
These numbers are similar to the case of world wealth inequality of 
households \cite{piketty2,jsp,complex}. 

Of course, the number of citations
characterizes only a part of scientific production of a university
and there are other important characteristics of university activity.
However, citations can be considered as a certain indication
of university success. Our results are based on the public database
of Google Scholar \cite{scholar}. It is possible that other citation databases,
e.g.  Web of Knowledge (Clarivate), would provide a more rigorous attribution
of scientist university affiliation. We argue that
our approach can bring
more detailed information when applied to
such more rigorous citation databases.
We also hope that the RJ thermalization concept can find applications 
in various other aspects of social phenomena and social physics
that attract growing interest of physicists \cite{socio}.


\section*{Appendix}
\setcounter{equation}{0}
\renewcommand{\theequation}{A\arabic{equation}}
\setcounter{figure}{0}
\renewcommand\thefigure{A\arabic{figure}}
\renewcommand\thesection{A\arabic{section}}
\renewcommand\thesubsection{A\arabic{subsection}}
\renewcommand{\figurename}{Appendix Figure}

\setcounter{figure}{0}
\renewcommand{\thefigure}{A\arabic{figure}} 

\subsection{Additional figures}

\begin{figure}[h]
\begin{center}
\includegraphics[width=0.95\columnwidth]{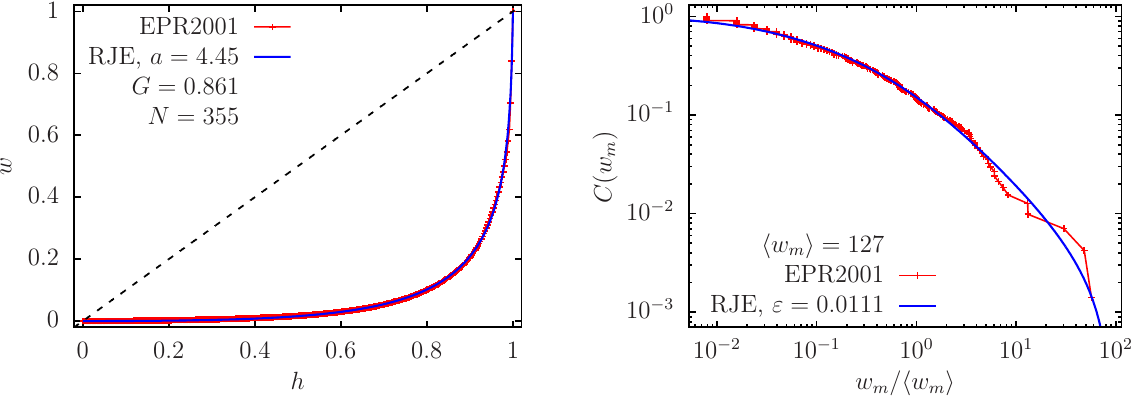}%
\end{center}
\caption{\label{figA1}
\label{fig_EPR2001_multi}
As Fig.~\ref{fig1} for the papers citing the original Einstein 
EPR paper during the year 2001. 
}
\end{figure}

Here we provide two additional figures related to Section 5. 
First, Fig.~\ref{figA1} shows Lorenz and Pareto curves in the same 
style as Fig.~\ref{fig1}, for the set of 355 papers published in 2001 
that cite the historical EPR 1937 paper of Einstein-Podolsky-Rosen 
\cite{epr} and where the wealth quantity corresponds to the number of 
citations which each of these papers has obtained since its publication. 
The average citation number is 127 and the 
Gini coefficient $G=0.861$ is rather high and 
comparable to the value of Einstein's publications.
The RJE curves, with $a=4.45$ and $\eps=0.0111$, 
agree very well with the data curves, with some small deviations at the 
tail of the Pareto distribution. 


\begin{figure}[h]
\begin{center}
\includegraphics[width=0.95\columnwidth]{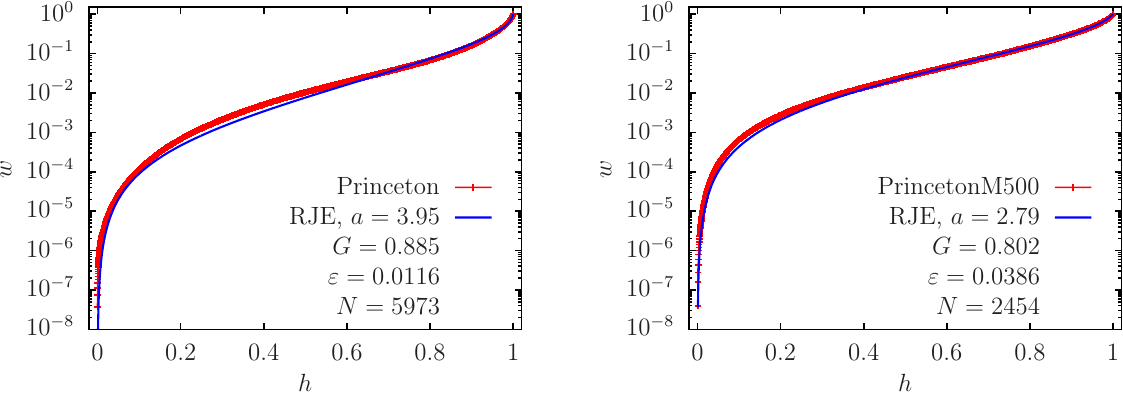}%
\end{center}
\caption{\label{figA2}
\label{fig_Princeton_lorlog_PrincetonM500_lorlog}
Comparison of data Lorenz curves (red data points) of Princeton with 
RJE Lorenz curves (blue) in logarithmic representation 
for the wealth variable $w$ where the left (right) panel corresponds 
to case where all affiliated scientists have been taken into account 
(only those with at least 500 citations indicated by the label ``M500''). 
For both cases all RJE 
parameters and $G$ are given in the figure. }
\end{figure}

In Fig.~\ref{figA2}, we illustrate for the example of Princeton the effect 
of the minimal threshold of 500 citations used for the data 
in Figs.~\ref{fig6}-\ref{fig10}. 
Both panels of Fig.~\ref{figA2} show Lorenz curves of Princeton 
affiliated scientists (in a logarithmic scale for the cumulated wealth 
$w(h)$). The left panel corresponds to the case of 5973 Princeton 
affiliated scientists with at least 1 citation and the right 
panel to the case of 2454 Princeton affiliated scientists with at least 
500 citations (same data as in Fig.~\ref{fig6}) 
and where the wealth corresponds to the reduced citation 
number (being citation number minus 500 for each scientist). 
In both cases, the optimal RJE curve is very close to the real data curve, 
even for small values of $h$ and in a logarithmic scale for $w$ but 
with smaller values of $a$ and $G$ for the latter case. This observation 
seems plausible since in the first case the poor phase (scientists with 
few citations) is strongly increased.

\subsection{Case of one integral of motion} 

In refs. \cite{jsp,rmtprl,ourfiber,complex}, a 
nonlinear oscillator system with complex amplitudes $C_m(t)$ 
of the form $i\frac{\partial C_m(t)}{\partial t}=E_m C_m(t)+\beta 
F_m(\{C_l(t)\})$ was considered where $F_m(\{C_l(t)\})$ is some complicated 
nonlinear function of all the amplitudes $C_l(t)$ obtained from 
some special potential (see, e.g., \cite{jsp} for details) and 
$\beta$ is a parameter for the strength of the nonlinear perturbation. 
The particular form of $F_m(\{C_l(t)\})$ used in the above references 
ensured that in this nonlinear oscillator system there are two exact 
integrals of motion which are 
$\mathcal{H}=\sum_m E_m |C_m(t)|^2+\beta V(\{C_l(t)\})$ and 
$\eta=\sum_m|C_m(t)|^2$ where $V(\{C_l(t)\})$ is the potential 
associated with the nonlinear term $F_m(\{C_l(t)\})$. 
Typically, once the system is thermalized (at long times) 
the contribution of the potential 
is small, $V\sim \beta/N$, and we have $\mathcal{H}\approx 
\sum_m E_m |C_m(t)|^2$ (see \cite{jsp} for some explanations 
concerning subtle deviations of this at small times). 

This type of system corresponds to a special type of microcanonical ensemble 
with fixed energy $\mathcal{H}=E$ and 
the additional constraint that $\eta$ is conserved (typically 
the value $\eta=1$ is chosen). According to standard techniques of 
statistical physics, such a system can be replaced (for $N\gg 1$) 
by a grand canonical ensemble. In this approach the amplitudes $C_m(t)$ have 
(in equilibrium) a Gaussian distribution with the RJ average 
$\rho_m=\langle |C_m(t)|^2\rangle =T/(E_m-\mu)$ where 
the temperature $T$ and the chemical potential $\mu$ are 
determined to satisfy both constraints (for given values of 
$\mathcal{H}=E$, $\eta=1$). Using this expression for $\rho_m$, 
the construction of the associated Lorenz curves for 
the RJS and RJE models is explained in Section 4. 

However, the conservation of $\eta$ is a specific property of a 
certain type of potential (see, e.g., \cite{jsp}) but 
for other more generic potentials $V$ 
(or nonlinear terms $F_m$ in the oscillator equations) we typically have 
only one integral of motion which is $\mathcal{H}$ while the 
quantity $\eta(t)=\sum_m |C_m(t)|^2$ is not conserved. Such a system 
corresponds the usual standard microcanonical ensemble (with only 
one energy constraint). Here the entropy is given as 
$S_B=\ln\Omega(E)$ where $\Omega(E)$ is 
the phase space volume of the energy shell fixed by $\mathcal{H}=E$ and 
which behaves here as $\Omega(E)\sim E^{N-1/2}$. Then the microcanonical 
temperature is given by $T=(dS/dE)^{-1}=E/(N-1/2)\approx E/N$. 
Here it is necessary that all $E_m>0$ in order 
to have a finite volume $\Omega(E)$ 
of the energy shell. For such a system, one can 
show that the marginal distribution of $C_m$ 
(obtained by integrating the full distribution over all other $C_l$ 
for $l\neq m$) is for $N\gg 1$ 
also Gaussian with average $\rho_m=\langle |C_m(t)|^2\rangle=T/E_m$ and 
the quantity $\eta(t)$ has the average 
$\langle \eta\rangle=T\sum_m 1/E_m$ which has some value ($>0$) depending on 
$T=E/N$ (and the choice of the spectrum $E_m$). 

Now, the cumulated household and wealth quantities in the 
usual construction of the Lorenz curve are given by: 
$h(m)=\sum_{l=0}^{m-1} \rho_l/\langle\eta\rangle$ 
and $w(m)=\sum_{l=0}^{m-1} (E_l/E)\rho_l$. We note that these 
expressions do not depend on the initial choice of $E$ or $T=E/N$ since 
the dependence on $T$ drops out in both ratios: 
$\rho_l/\langle\eta\rangle=(1/E_l)/\sum_{l'} (1/E_{l'})$ and 
$E_l\rho_l/E=T/E=1/N$. Therefore, we choose for convenience $T$ (and $E=NT$) 
such that $\langle\eta\rangle =1$ resulting in the usual 
cumulated household expression $h(m)=\sum_{l=0}^{m-1} \rho_l$. 
On the other hand the Lorenz curves depend quite strongly on the 
value $\Delta_0=E_0>0$ of the smallest oscillator mode. 
We can extend the RJS and RJE models to 
the case of one integral (OI) called RJS OI and RJE OI models 
by choosing a spectrum $E_m=\Delta_0+m/N$ (RJS OI) 
or $E_m=\Delta_0+(e^{am/N}-1)/(e^a-1)$ (RJE OI with parameter $a$). 
In both cases, the bandwidth is $B=E_{N-1}-\Delta_0\approx 1$ and 
the initial level spacing $\Delta=E_1-\Delta_0$ is 
either $\Delta=1/N$ (RJS OI) or $\Delta=a/[N(e^a-1)]$ (RJE OI). Note that this 
convention for the RJE bandwidth is similar to the continuous case 
discussed in Appendix section 2 of \cite{complex} but it is different from 
\cite{jsp} where $B=(e^a-1)/a$ was used. Of course, 
the precise choice of $B$ does not affect the form 
of the Lorenz curves which are invariant with respect to global energy 
rescaling. 

\begin{figure}[htbp]
\begin{center}
\includegraphics[width=0.9\textwidth]{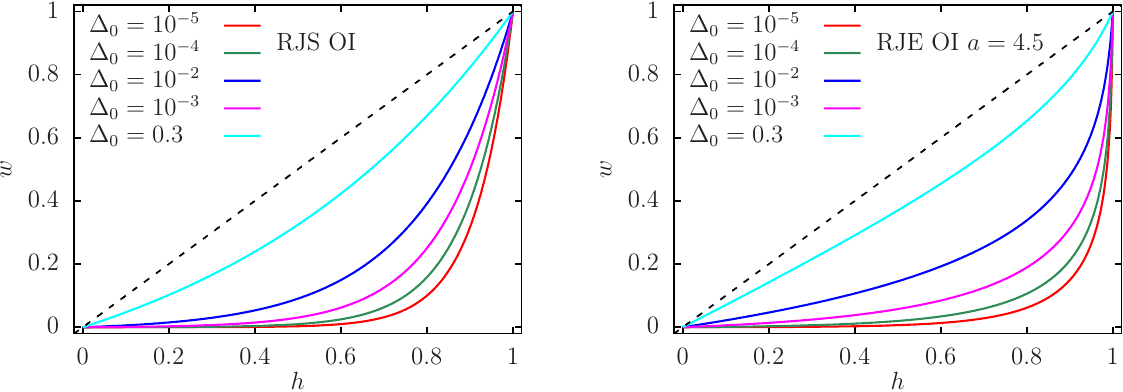}
\caption{\label{figA3}
\label{fig_oneint_gen}
Some Lorenz curves for the RJS OI model (left panel) and 
the RJE OI model (right panel) at $a=4.5$ (both for the 
continuous model with $N\to\infty$ for the number of 
oscillator modes and a bandwidth $B=1$; 
see text for details). 
The dashed black line corresponds to the case of perfect equipartition 
$w=h$. 
The different (full) curves correspond to different values of the 
ground state energy $\Delta_0=10^{-5},10^{-4},10^{-3},10^{-2},0.3$ 
(bottom to top). The respective values of the rescaled energy are 
$\eps=0.0868,0.108,0.144,0.207,0.382$
(for $a=0$ in the left panel)
and $\eps=0.00719,0.0105,0.0181,0.0382,0.128$
(for $a=4.5$ in the right panel) and 
the respective values of the Gini coefficients are 
$G=0.826,0.783,0.713,0.587,0.236$ (left) and 
$G=0.9,0.855,0.765,0.59,0.197$ (right). 
}
\end{center}
\end{figure}

In Fig.~\ref{figA3}, we show for the RJS OI and the RJE OI model at $a=4.5$ 
several Lorenz curves for the ground state energies 
$\Delta_0=10^{-5},10^{-4},10^{-3},10^{-2},0.3$. 
For each value of 
$\Delta_0$ and $a$, we have a corresponding value of the rescaled energy 
$\eps=(E-\Delta_0)/(E_{N-1}-\Delta_0)\approx E-\Delta_0=\sum_m (E_m-\Delta_0)\rho_m$ (since 
$E_{N-1}-\Delta_0\approx 1$ and where $E$ is fixed by the choice of 
$\langle\eta\rangle=1$) and the Gini coefficient $G$ (both given 
in the caption of Fig.~\ref{figA3} for all curves).
As for the RJE (two integral) case (see, e.g., 
Fig.~\ref{fig4} of \cite{complex}),
we see that with increasing $\eps$ we have 
decreasing values of $G$. However, in contrast to the RJE case, 
for the larger $\eps$ values (smaller $G$ values) there 
is a finite linear slope $w'(0)\approx \Delta_0/E=\Delta_0/(\Delta_0+\eps)$ 
which is an effect of the non-vanishing value of $\Delta_0$. The curves 
of Fig.~\ref{figA3} have been obtained for the formal 
continuous limit $N\to\infty$ 
using analytical expressions of $w(h)$ similar to those of \cite{complex} 
for the two integral case (see also below for the extension of 
this to the RJE OI case). However, we have also 
verified that the case of 
finite but large $N$ (e.g. $N=10^4$ or $N=10^5$) provides similar curves 
with similar values of $\eps$ and $G$ 
except for the two smallest $\Delta_0$ values for the RJS case 
where $\Delta_0$ is smaller or 
comparable to the initial level spacing $\Delta=1/N$ (for $N=10^4$).

The case of the RJE OI model with $\Delta_0>0$, can be associated with the 
usual RJE model with spectrum $\tilde E_m =E_m-\Delta_0=(e^{am/N}-1)/(e^a-1)$ 
and chemical potential $\mu=-\Delta_0$. Then, the values 
of $\rho_m=T/E_m=T/(\Delta_0+\tilde E_m)=T/(\tilde E_m-\mu)$ 
are identical between 
both models resulting in identical household quantities $h(m)$. 
However, the average energy $E$ for the RJE OI model corresponds to 
$E=\Delta_0+\tilde E$ where $\tilde E$ is the energy of the associated 
usual RJE model and we can establish the following relation between the 
cumulated wealth variables of both models:
\begin{align}
\nonumber
w^{\rm (RJE\ OI)}(h(m)) &
=\frac{T}{E}\sum_{l=0}^{m-1} \frac{E_l}{E_l}=\frac{Tm}{E}\quad,\\
\nonumber
w^{\rm (RJE)}(h(m))&
=\frac{T}{\tilde E}\sum_{l=0}^{m-1} \frac{\tilde E_l}{\tilde E_l-\mu}
=\frac{T}{\tilde E}\sum_{l=0}^{m-1} \left(1+\frac{\mu}{\tilde E_l-\mu}\right)
=\frac{Tm}{\tilde E}+\frac{\mu}{\tilde E}h(m)\folgt\\
\label{LorOI}
w^{\rm (RJE)}(h)&=\frac{E}{E-\Delta_0}w^{\rm (RJE\ OI)}(h) 
-\frac{\Delta_0}{E-\Delta_0}h
\end{align}
since $\mu=-\Delta_0$ and $\tilde E=E-\Delta_0$. 
In (\ref{LorOI}) the linear term in $h$ compensates the 
finite slope ${w'}^{\rm (RJE\ OI)}(0)=\Delta_0/E$ such 
that ${w'}^{\rm (RJE)}(0)=0$ which is the typical behavior of 
the usual RJE model (since $\tilde E_0=0$). 

For completeness, we remind some of the analytical results for the 
continuous 
limit $N\to\infty$ of \cite{complex}. In this approach, the sums defining 
$h(m)$ and $w(m)$ are replaced by integrals over a rescaled index variable 
$s\in[0,1]$ corresponding to $s=m/N$ (or $s=l/N$), which gives a closed 
(somewhat complicated) expression $h(s)$ depending on $a$ and $\mu$ (for 
the usual RJE case; see eq. (A11) of \cite{complex}). 
Then $\mu$ is determined by the implicit equation 
$h(s=1)=1$ which gives
\begin{align}
\label{norm1intC}
1&=\frac{\eps-\mu}{a(D^{-1}(a)+\mu)}\ln\left(
\frac{ e^{-a}(1-\mu) }{-\mu}\right)
\folgt\eps=\mu+\frac{a(D^{-1}(a)+\mu)}
{\ln\left(
\frac{ e^{-a}(1-\mu) }{-\mu}\right)}\ .
\end{align}
Inverting 
the expression for $h=h(s)$, one can determine the dependence $s=s(h)$ 
of the rescaled index $s$ on $h$ which gives \cite{complex}
\begin{align}
\label{eqsofh}
s(h)=-\frac{1}{a}\ln\left\{1+\frac{\mu D(a)}
{1+\mu D(a)}
\left[\exp\left(\frac{a(1+\mu D(a))h}{D(a)(\eps-\mu)}\right)-1\right]
\right\}\quad,\quad D(a)=e^a-1\ .
\end{align}
For the RJE case the cumulated wealth fraction is given by 
$w^{\rm (RJE)}(h)=(1-\mu/\eps)s(h)+(\mu/\eps)h$ \cite{complex} 
using the above expression for $s(h)$. Furthermore, there are also 
analytic expressions for the Gini coefficient $G$ and the Pareto function 
$C^{\rm (RJE)}(w_m)$ which can be found in \cite{complex} 
(see Eqs. (A20), (A24-A26) and (A32) there). 
These results were obtained for the RJE case (with two integrals of 
motion). However, they can be easily translated to the RJE OI case. For this, 
we need to replace $\mu\to-\Delta_0$ and use the 
2nd expression of (\ref{norm1intC}) which allows to determine directly 
$\eps$ for a given value of $\Delta_0$ (note for the RJE case one 
provides $\eps$ and solves the 1st expression of (\ref{norm1intC}) 
numerically to determine $\mu$). 

However, the cumulated wealth variable for the RJE OI case is now given 
by the simpler expression $w^{\rm (RJE\ OI)}=s(h)$ using again (\ref{eqsofh}) 
but with the replacement $\mu\to -\Delta_0$. Note that both wealth expressions 
for RJE and RJE OI are in agreement with (\ref{LorOI}) and 
using the identifications $E-\Delta_0=\eps$ and $\Delta_0=-\mu$. The Gini coefficient expression 
(A20) of \cite{complex} for the RJE case, is now modified for the RJE OI 
case according to
\begin{align}
G^{\rm (RJE\ OI)} &=1-\frac{2}{a}F(x,y)\label{eqG1int}
\end{align}
where the quantity $F(x,y)$ is the same as in \cite{complex} 
(see (A24-A26) there). Finally, the RJE OI Pareto function is 
obtained by a simple argument shift from the RJE Pareto function, 
i.e. $C^{\rm (RJE\ OI)}(w_m)=1$ for $w_m<\Delta_0$ and 
$C^{\rm (RJE\ OI)}(w_m)=C^{\rm (RJE)}(w_m-\Delta_0)$ for 
$w_m\ge \Delta_0$ (see (A32) of \cite{complex} for $C^{\rm (RJE)}(w_m)$).

\begin{figure}[h]
\begin{center}
\includegraphics[width=0.95\columnwidth]{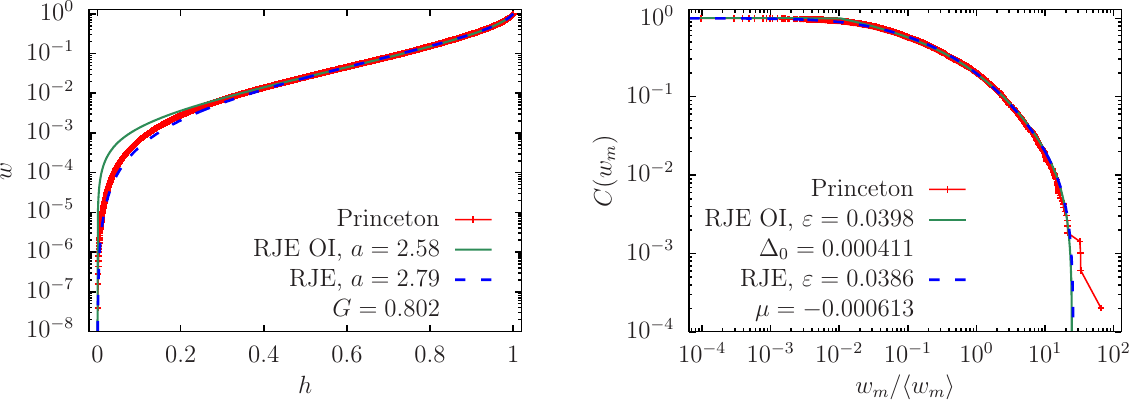}%
\end{center}
\caption{\label{figA4}
\label{fig_PrincetonM500_oneint}
{\em Left:} Comparison of the citation Lorenz curve for Princeton 
with citation threshold 500 (red data curve) with both 
theoretical RJE OI model (full green curve) and RJE model 
(dashed blue curve). 
The $x$-axis corresponds to the 
cumulative fraction of households/affiliated scientists ($h$) and the $y$-axis 
(in logarithmic scale) to the cumulative fraction of wealth/reduced 
citations ($w$). 
The parameters of $a$, $\eps$, $\mu$ or $\Delta_0$ 
have been determined such that the 
theoretical curves have the same Gini coefficient $G=0.802$ as the 
red data curve and such that the geometrical curve distance between 
theoretical curves and real data curve is minimized. 
{\em Right:} Pareto curve $C(w_m)$ for the same data (red data curve) 
where $C(w_m)$ represents the fraction of Princeton affiliated scientists 
(with at least 500 citations) having more reduced citations 
then than $w_m$ (analogous to wealth). 
The $x$-axis corresponds to rescaled 
values $w_m/\langle w_m\rangle$ where 
$\langle w_m\rangle$ is the average reduced citation number per scientist. 
The full green and dashed blue curve correspond to the 
theoretical RJE OI or RJE Pareto curve with the same parameters as in the left 
panels (all parameters are given in the figure). 
}
\end{figure}

For small values $\Delta_0\ll E$ the relation (\ref{LorOI}) shows 
that the Lorenz curves between both models are very similar. 
We have, for a few examples of real data, computed optimal parameters 
$\Delta_0$, $a$ (and the related value of $\eps$) such that the 
real data Lorenz curve fits in the best possible way (minimizing the 
geometrical curve distance) the theoretical RJE OI Lorenz curve and that 
both curves have the same Gini coefficient $G$. 
In Fig.~\ref{figA4}, we compare the data of Princeton used in 
Fig.~\ref{fig6} with a threshold of at least 500 citations, 
with the optimal RJE OI Lorenz and Pareto curves (full green lines). 
For illustration, also the optimal RJE curves (dashed blue lines) are shown. 
For the Lorenz curve a logarithmic representation for $w(h)$ is used 
in Fig.~\ref{figA4}. The optimal RJE OI curve is slightly different 
from the optimal RJE curve and data at small values of $h$ but apart from 
this both theoretical curves are very close. In a normal representation 
(without logarithmic scale for $w(h)$ and not shown here), 
both theoretical Lorenz curves are identical on graphical precision. 
The difference at small $h$ is due to the (small) finite initial slope of the 
RJE OI model with $w'(0)=\Delta_0/(\Delta_0+\eps)$ 
while for the RJE case and also the real Princeton data 
we have $w'(0)=0$. However, since $\Delta_0\ll \eps$ for the parameters 
shown in Fig.~\ref{figA4}, this effect is very small. 

\begin{figure}[h]
\begin{center}
\includegraphics[width=0.95\columnwidth]{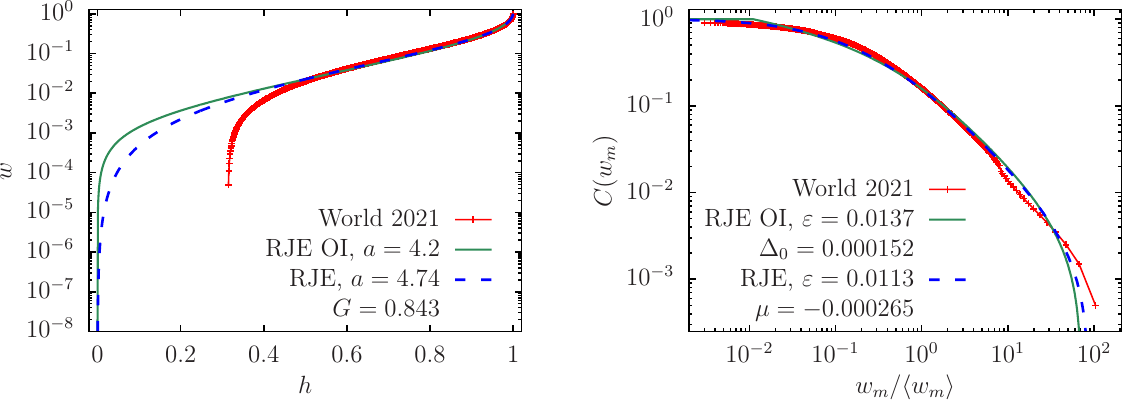}%
\end{center}
\caption{\label{figA5}
\label{fig_World_2021_oneint}
As Fig.~\ref{figA4}, but using World 2021 wealth data 
from \cite{piketty2}. Note that the absent red data points for 
$h<0.3$ in the left panel are due to the logarithmic scale and the 
slightly negative values of $w(h)$ in this interval 
which are due to a situation of debt for these households. 
}
\end{figure} 

Fig.~\ref{figA5} shows a similar comparison for the wealth data of 
World 2021 \cite{piketty2} which was also analyzed in \cite{complex}. 
Here, the agreement of all cases is similar to the Princeton data. 
However, we mention that in the World 2021 data there are small negative 
cumulated wealth values for $h\le 0.3$ (situation of debt) 
implying that there are no visible Lorenz curve data points for this 
interval due to the logarithmic representation.

\begin{figure}[h]
\begin{center}
\includegraphics[width=0.95\columnwidth]{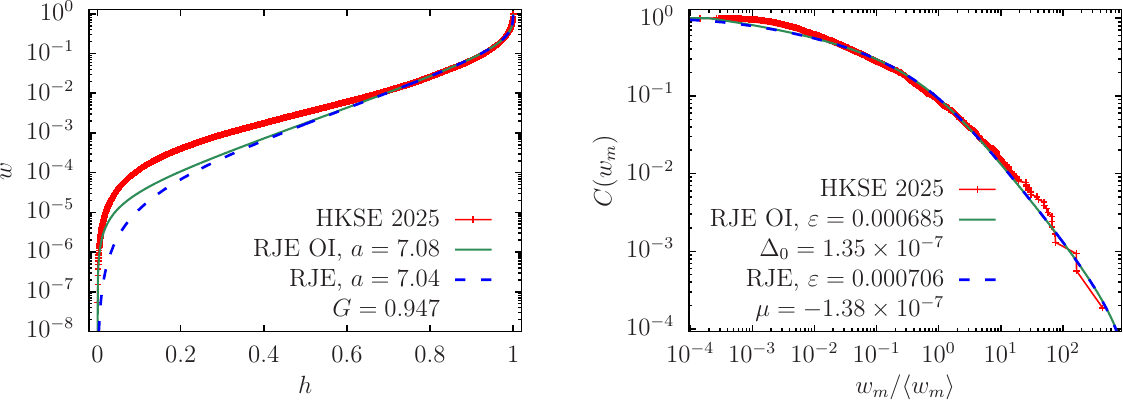}%
\end{center}
\caption{\label{figA6}
\label{fig_HKSE_2025_oneint}
As Fig.~\ref{figA4}, but using Hong Kong 2025 stock exchange data 
(see also \cite{complex}) where the wealth variable corresponds to the 
market capitalization of companies. Note that the RJE OI and RJE Pareto 
curves are essentially identical (on graphical precision) for 
most values of $w_m$ (in particular at the tail). 
}
\end{figure}

Finally, in Fig.~\ref{figA6} we show a similar pair of figures for the 
Hong Kong stock exchange of 2025 (here wealth = company market capitalization) 
using data already analyzed in \cite{complex}. Here, the difference between 
the two theoretical curves is very small due to the particularly small 
value of $\Delta_0$ and in particular the two Pareto curves 
are essentially identical (except a bit for small $w_m$). 
Also the agreement, between the theoretical curves and real HKSE 2025 data 
is very good. There are some visible deviations at small $h$ for the 
Lorenz curves but these are mostly an effect of the logarithmic representation 
and appear only for small wealth values $w(h)<10^{-2}$ which are essentially 
not visible in normal representation. 

\subsection{Lorenz and Pareto curves of poker tournaments} 

\begin{figure}[h]
\begin{center}
  \includegraphics[width=0.9\columnwidth]{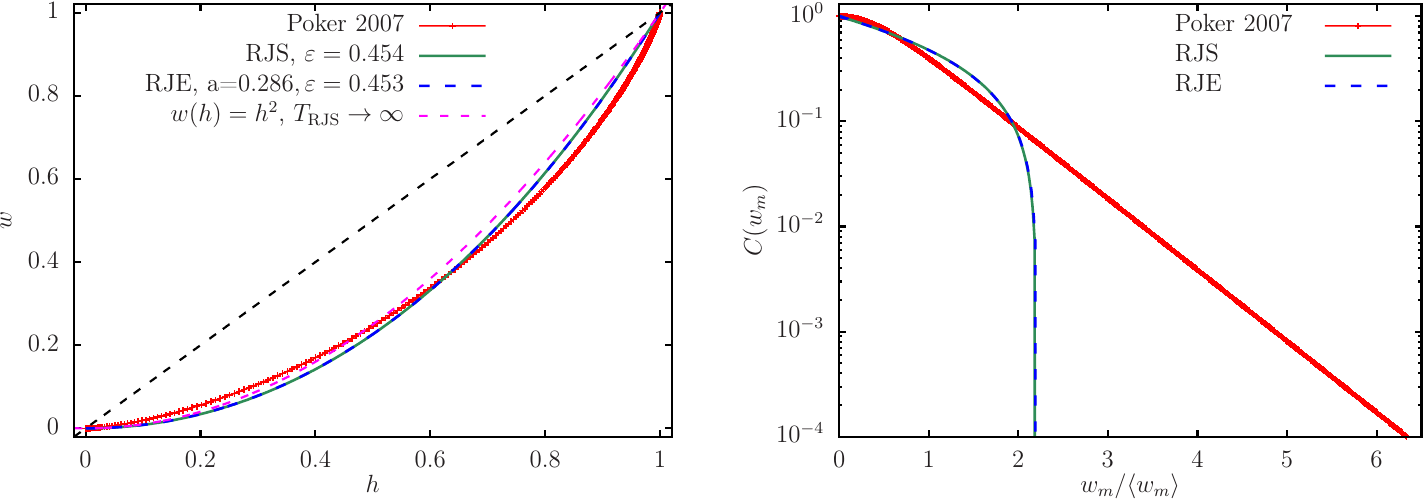}%
\end{center}
\caption{\label{figA7}
{\em Left:} Lorenz curve for poker tournaments
with Gini coefficient $G=0.365$ (red symbols) \cite{poker1,poker2},
RJS and optimal RJE theory curves at same $G$ are shown by full green 
and dashed blue curves that practically coincide. The dashed magenta curve 
shows the high temperature limit for the RJS model $w=h^2$ corresponding 
to $\rho_m=$const. with $G=1/3$ and the dashed black straight line corresponds 
to the case of perfect equipartition $w=h$ \cite{jsp}. 
RJS and RJE parameters $\eps$, $a$ are given in the left panel.
{\em Right:} Pareto curves $C(w_m)$ versus $w_m/\langle w_m\rangle$ 
for the same data and theoretical RJS and RJE curves for the same 
parameters as in the left panel. In contrast to other Pareto figures there 
is only a logarithmic scale for the $y$-axis but not for the $x$-axis 
to be coherent with figures from \cite{poker1}. 
}
\end{figure}

Here in Fig.~\ref{figA7} we show results for Lorenz and Pareto curves for 
poker tournaments data from \cite{poker1,poker2} and the 
corresponding theoretical curves for the RJS and RJE model (with optimal 
parameter $a$; see discussion at the end of Section \ref{sec6}).

\authorcontributions{All authors equally contributed to all stages of this work.
}

\acknowledgments{We thank C.Sire for discussions of his results of poker tournaments
  and providing us with the related numerical data \cite{poker1,poker2}. 
}
  
\funding{The authors acknowledge support from 
NANOX $N^\circ$ ANR-17-EURE-0009 in the framework of 
the Programme Investissements d'Avenir (project MTDINA).
}


\conflictsofinterest{The authors declare no conflict of interest.
} 



\reftitle{References}

\newpage
\setcounter{figure}{0}
\renewcommand\thefigure{S\arabic{figure}}
\setcounter{equation}{0}
\renewcommand\theequation{S\arabic{equation}}
\renewcommand{\figurename}{SupMat Figure}

\section*{
Supplementary Material for \\ 
\vskip 0.2cm
\noindent Gini coefficient of scientists and universities}
 
 \bigskip

\noindent by
K.~M.~Frahm $^1$ and D.~L.~Shepelyansky $^1$\\
\begin{description}
\item[$^1$] Univ Toulouse, CNRS, Laboratoire de Physique Th\'eorique,  
Toulouse, France
\end{description}
\bigskip

Below figures in the same style as Fig.~1 for 
Lorenz and Pareto curves 
are shown for additional examples concerning 20 scientists: 
Albert, Barabasi, Devoret, Doudna, Frahm, Grebogi, Haroche, Hopfield, Ketterle, Klitzing, Myerson, Newman, Parisi, Pikovsky, Shepelyansky, Sire, Thouless, Wineland, Zeilinger, Zoller (SupMat Figures S1-S20) 
and 3 universities: ESPCI, ETH, Technion (SupMat Figures S21-S23) 
with citation threshold of 500 for the university cases. 
In all cases, the resulting optimal RJE parameters, $a$, $\eps$, the 
Gini coefficient $G$, number $N$ of papers/affiliated scientists (with at 
least 500 citations) and the values of the average (reduced) citation number 
$\langle w_m\rangle$ are given in the figures. Additional parameters 
for these cases are given Tables 1 or 2.

\begin{figure}[h]
\begin{center}
\includegraphics[width=0.95\columnwidth]{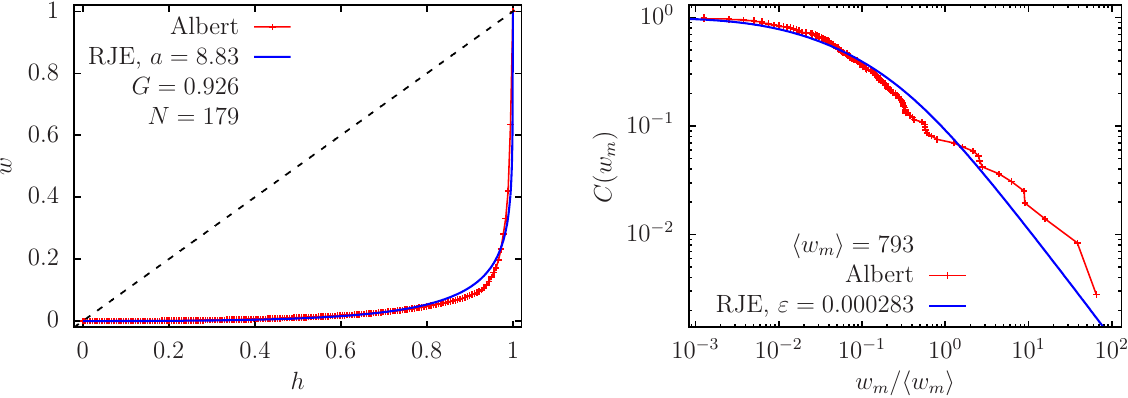}%
\end{center}
\caption{\label{figS1}
\label{fig_albert1_multi}
As Fig.~\ref{fig1} for Albert.
}
\end{figure}

\begin{figure}[h]
\begin{center}
\includegraphics[width=0.95\columnwidth]{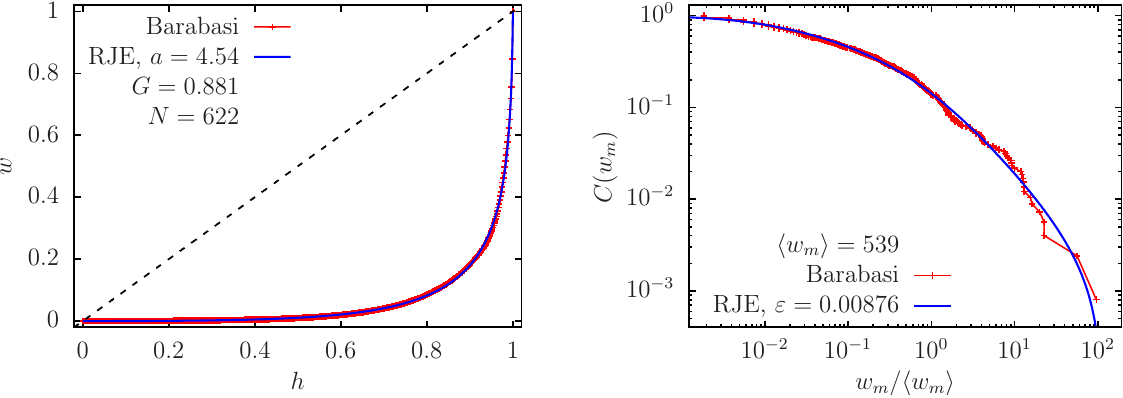}%
\end{center}
\caption{\label{figS2}
\label{fig_barabasi_multi}
As Fig.~\ref{fig1} for Barabasi.
}
\end{figure}

\begin{figure}[h]
\begin{center}
\includegraphics[width=0.95\columnwidth]{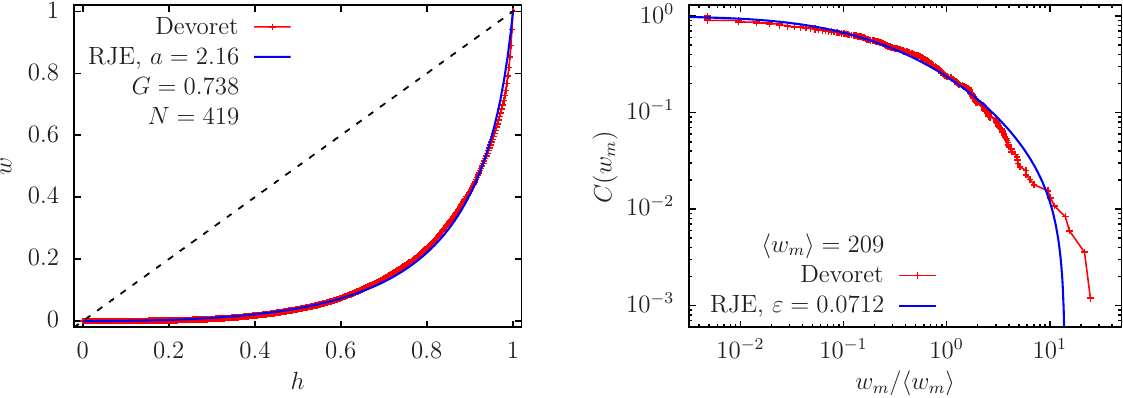}%
\end{center}
\caption{\label{figS3}
\label{fig_devoret1_multi}
As Fig.~\ref{fig1} for Devoret.
}
\end{figure}

\begin{figure}[h]
\begin{center}
\includegraphics[width=0.95\columnwidth]{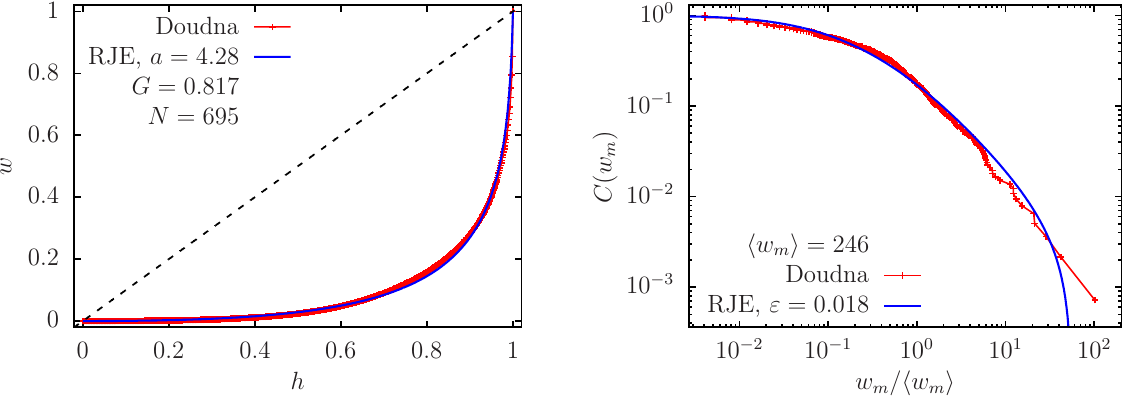}%
\end{center}
\caption{\label{figS4}
\label{fig_doudna1_multi}
As Fig.~\ref{fig1} for Doudna.
}
\end{figure}

\begin{figure}[h]
\begin{center}
\includegraphics[width=0.95\columnwidth]{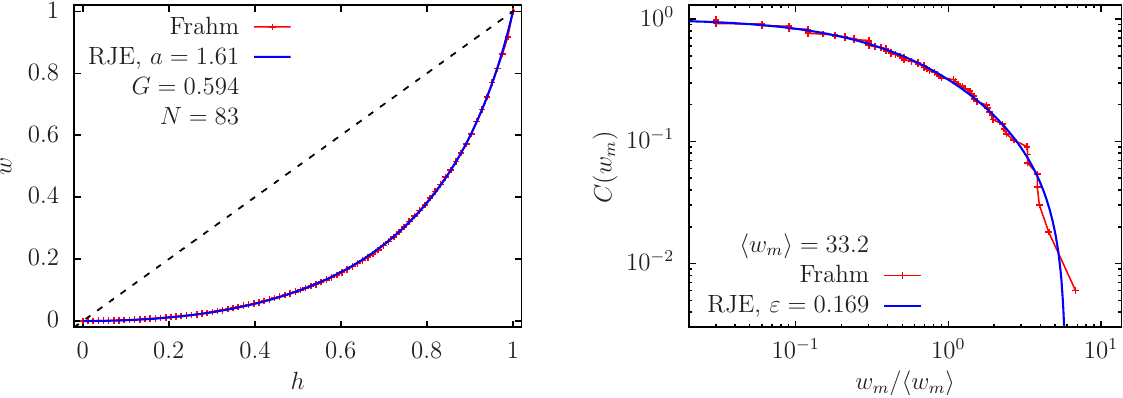}%
\end{center}
\caption{\label{figS5}
\label{fig_frahm1_multi}
As Fig.~\ref{fig1} for Frahm.
}
\end{figure}

\begin{figure}[h]
\begin{center}
\includegraphics[width=0.95\columnwidth]{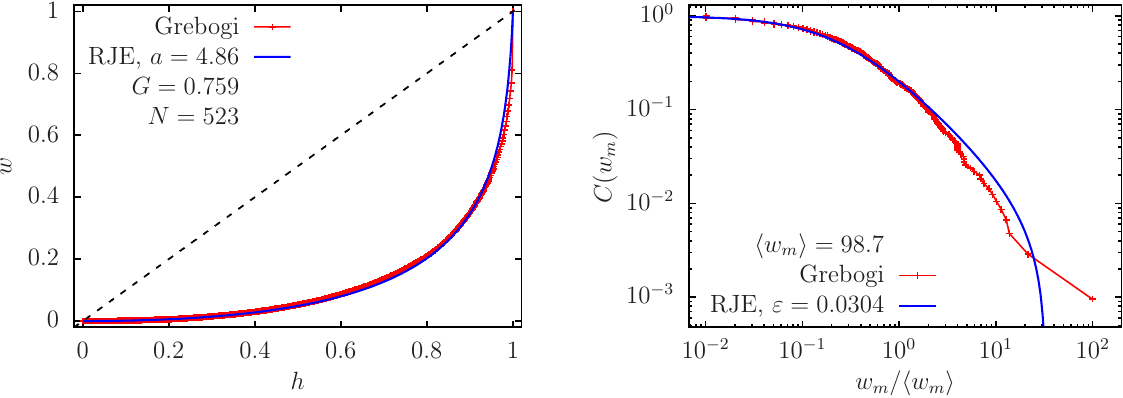}%
\end{center}
\caption{\label{figS6}
\label{fig_grebogi1_multi}
As Fig.~\ref{fig1} for Grebogi.
}
\end{figure}

\begin{figure}[h]
\begin{center}
\includegraphics[width=0.95\columnwidth]{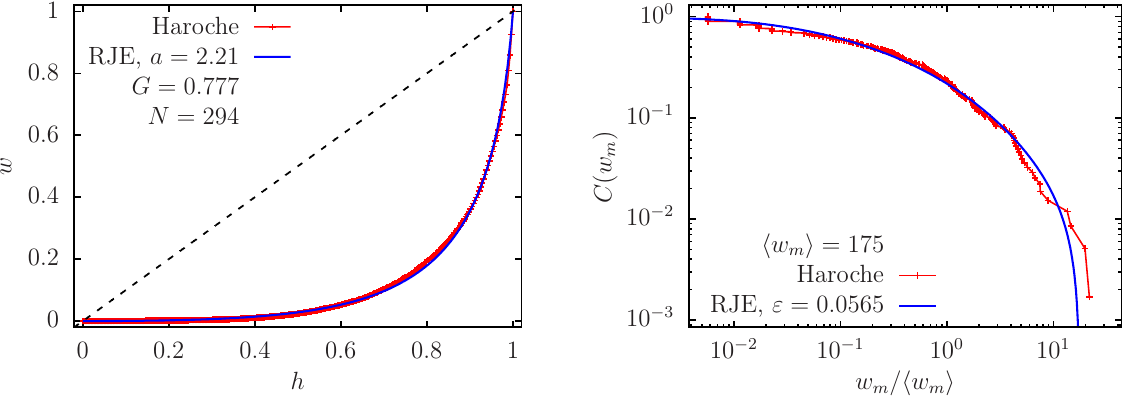}%
\end{center}
\caption{\label{figS7}
\label{fig_haroche1_multi}
As Fig.~\ref{fig1} for Haroche.
}
\end{figure}

\begin{figure}[h]
\begin{center}
\includegraphics[width=0.95\columnwidth]{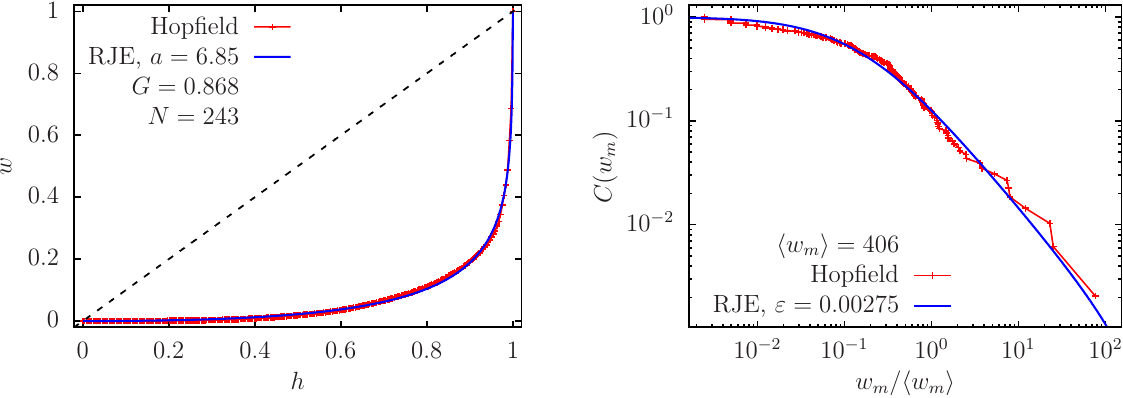}%
\end{center}
\caption{\label{figS8}
\label{fig_hopfield1_multi}
As Fig.~\ref{fig1} for Hopfield.
}
\end{figure}

\begin{figure}[h]
\begin{center}
\includegraphics[width=0.95\columnwidth]{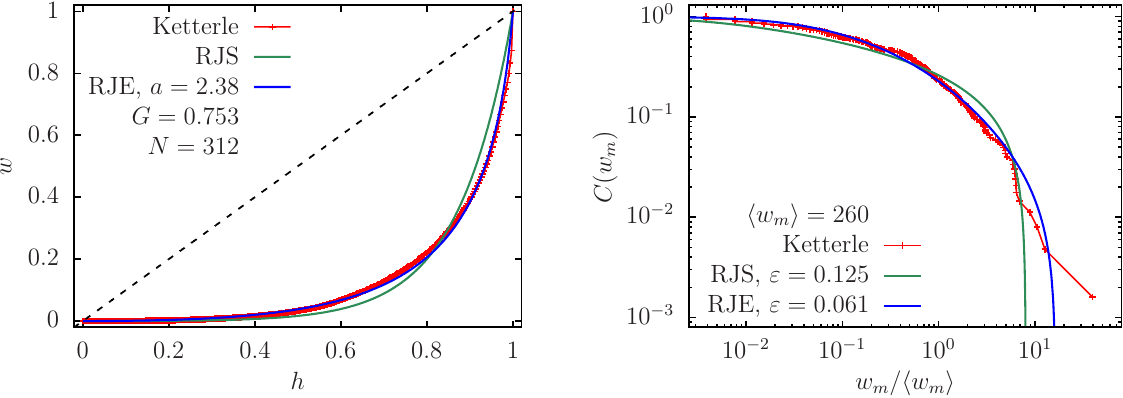}%
\end{center}
\caption{\label{figS9}
\label{fig_ketterle1_multi}
As Fig.~\ref{fig1} for Ketterle.
}
\end{figure}

\begin{figure}[h]
\begin{center}
\includegraphics[width=0.95\columnwidth]{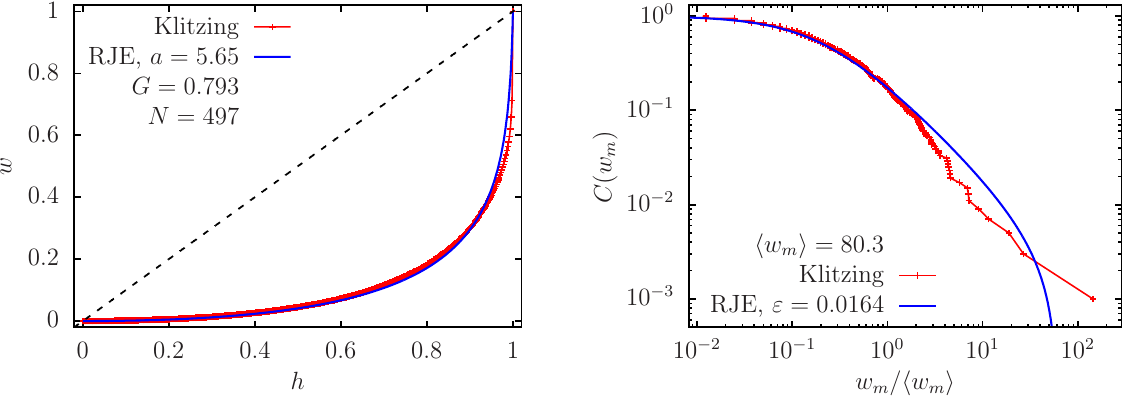}%
\end{center}
\caption{\label{figS10}
\label{fig_klitzing1_multi}
As Fig.~\ref{fig1} for Klitzing.
}
\end{figure}

\begin{figure}[h]
\begin{center}
\includegraphics[width=0.95\columnwidth]{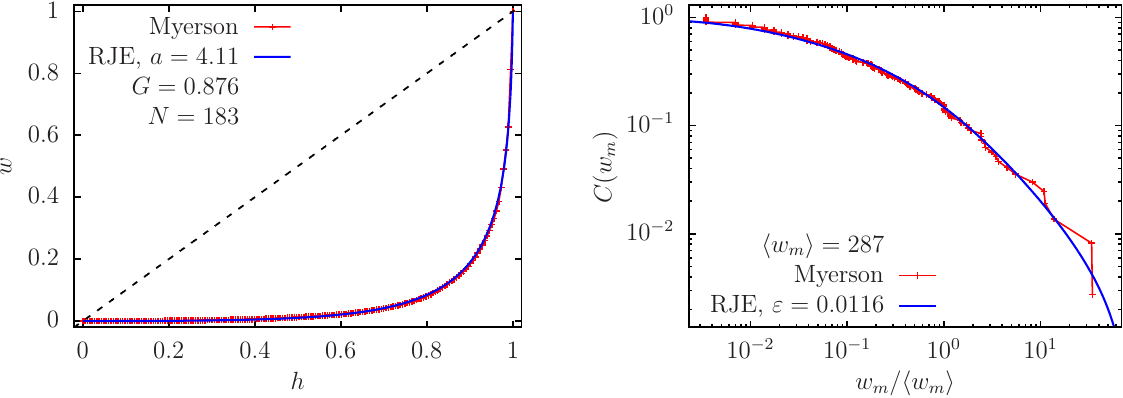}%
\end{center}
\caption{\label{figS11}
\label{fig_myerson1_multi}
As Fig.~\ref{fig1} for Myerson.
}
\end{figure}

\begin{figure}[h]
\begin{center}
\includegraphics[width=0.95\columnwidth]{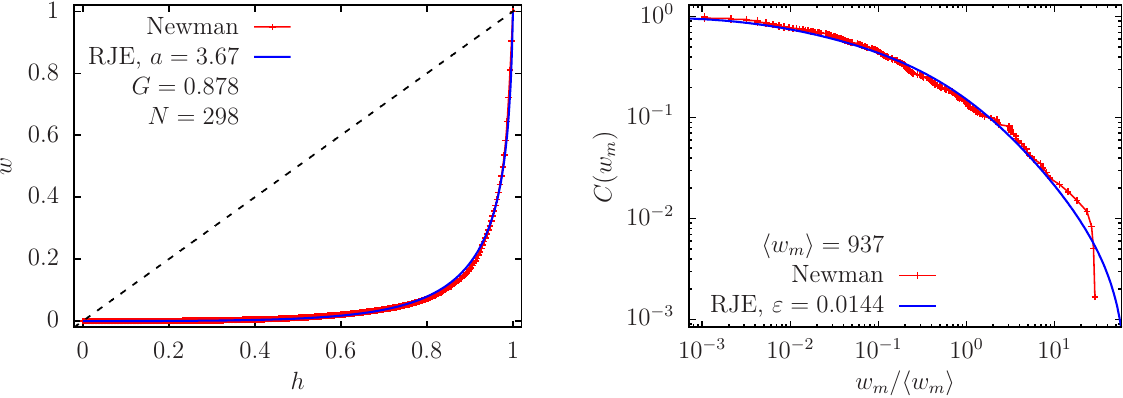}%
\end{center}
\caption{\label{figS12}
\label{fig_newman1_multi}
As Fig.~\ref{fig1} for Newman.
}
\end{figure}

\begin{figure}[h]
\begin{center}
\includegraphics[width=0.95\columnwidth]{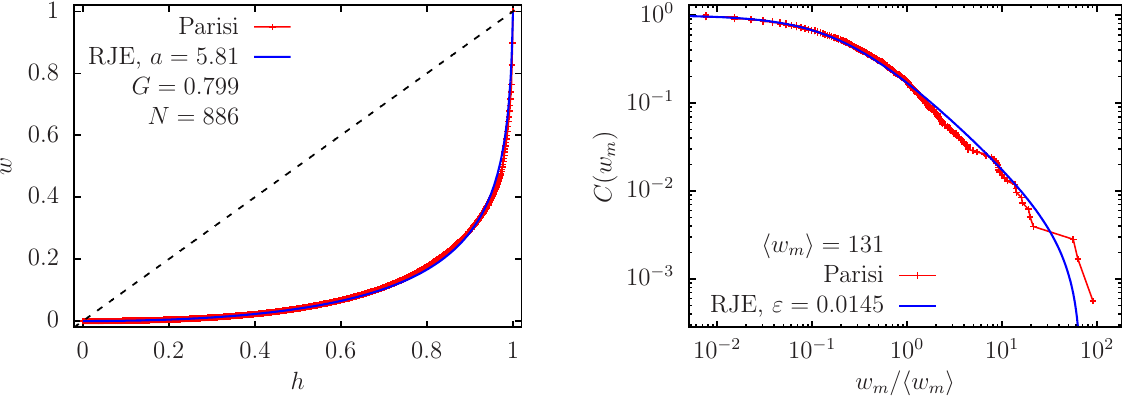}%
\end{center}
\caption{\label{figS13}
\label{fig_parisi1_multi}
As Fig.~\ref{fig1} for Parisi.
}
\end{figure}

\begin{figure}[h]
\begin{center}
\includegraphics[width=0.95\columnwidth]{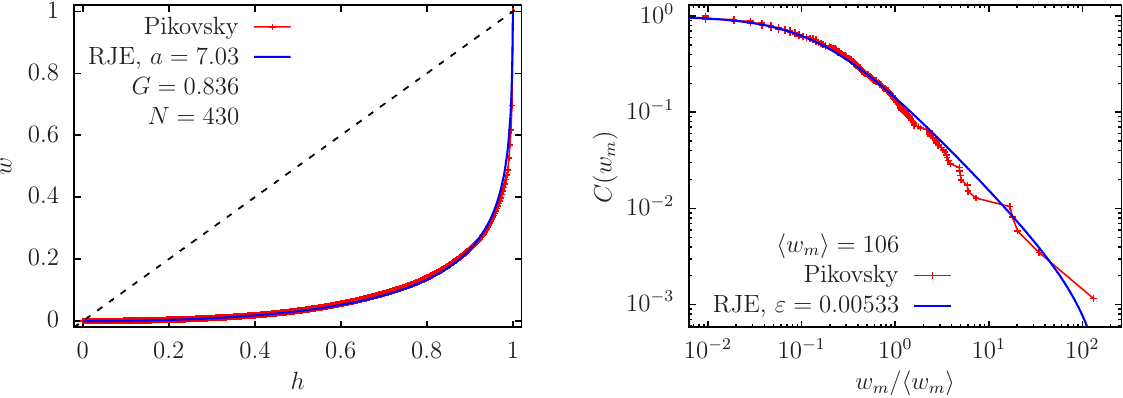}%
\end{center}
\caption{\label{figS14}
\label{fig_pikovsky1_multi}
As Fig.~\ref{fig1} for Pikovsky.
}
\end{figure}

\begin{figure}[h]
\begin{center}
\includegraphics[width=0.95\columnwidth]{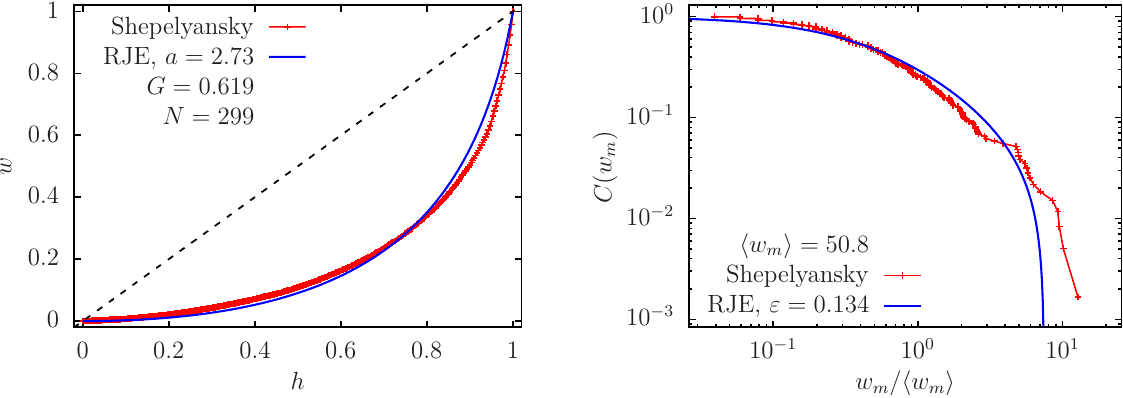}%
\end{center}
\caption{\label{figS15}
\label{fig_shepelyansky1_multi}
As Fig.~\ref{fig1} for Shepelyansky.
}
\end{figure}

\begin{figure}[h]
\begin{center}
\includegraphics[width=0.95\columnwidth]{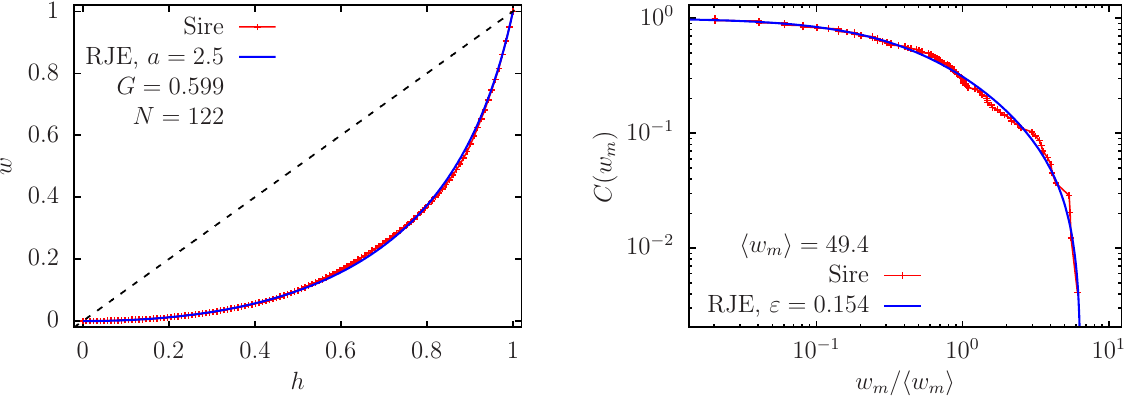}%
\end{center}
\caption{\label{figS16}
\label{fig_sire1_multi}
As Fig.~\ref{fig1} for Sire.
}
\end{figure}

\begin{figure}[h]
\begin{center}
\includegraphics[width=0.95\columnwidth]{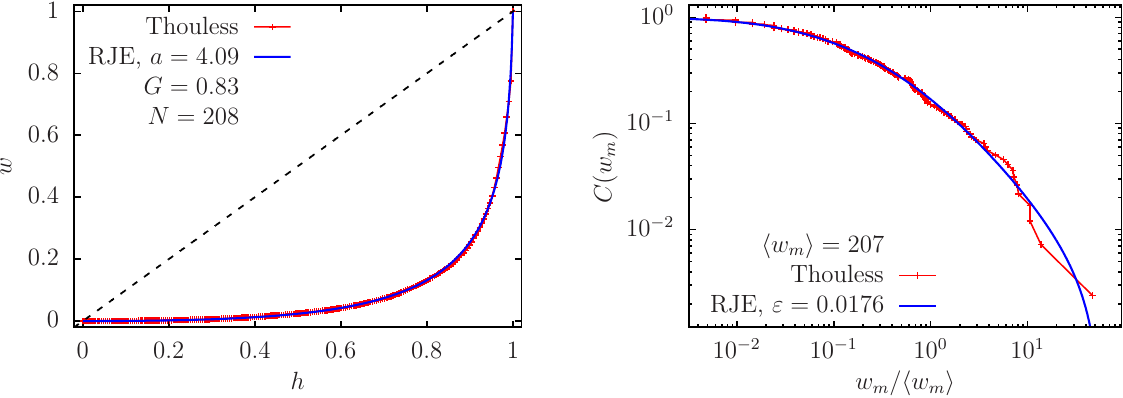}%
\end{center}
\caption{\label{figS17}
\label{fig_thouless1_multi}
As Fig.~\ref{fig1} for Thouless.
}
\end{figure}

\begin{figure}[h]
\begin{center}
\includegraphics[width=0.95\columnwidth]{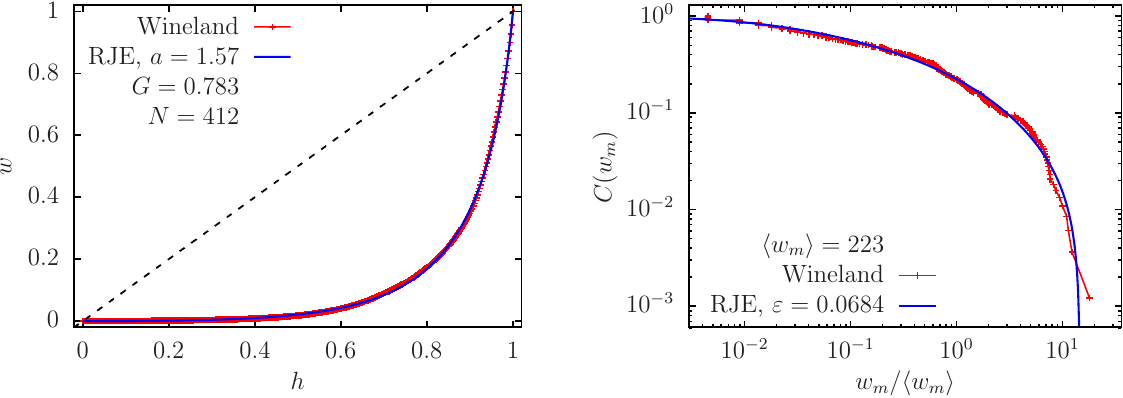}%
\end{center}
\caption{\label{figS18}
\label{fig_wineland1_multi}
As Fig.~\ref{fig1} for Wineland.
}
\end{figure}

\begin{figure}[h]
\begin{center}
\includegraphics[width=0.95\columnwidth]{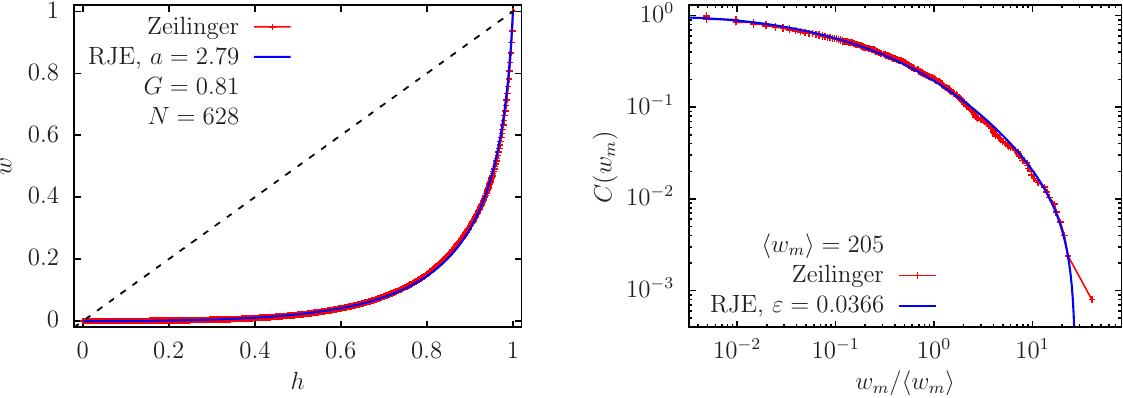}%
\end{center}
\caption{\label{figS19}
\label{fig_zeilinger1_multi}
As Fig.~\ref{fig1} for Zeilinger.
}
\end{figure}

\begin{figure}[h]
\begin{center}
\includegraphics[width=0.95\columnwidth]{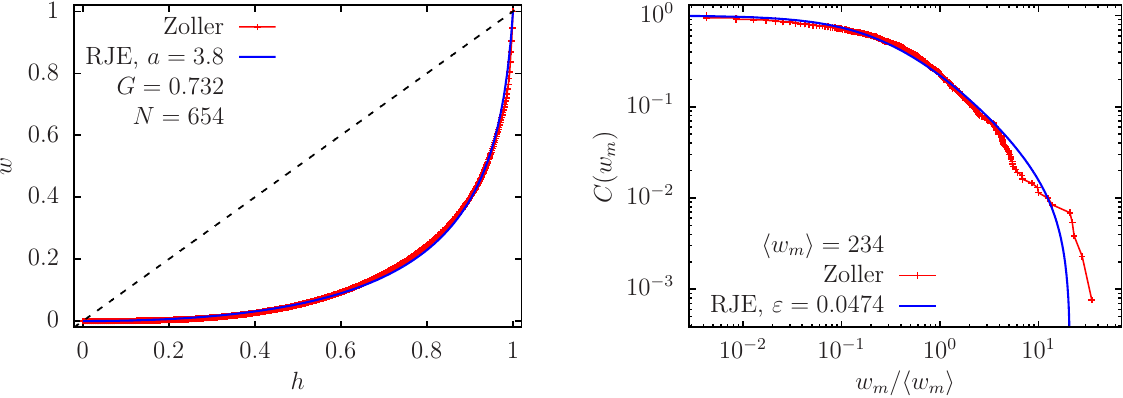}%
\end{center}
\caption{\label{figS20}
\label{fig_zoller1_multi}
As Fig.~\ref{fig1} for Zoller.
}
\end{figure}

\begin{figure}[h]
\begin{center}
\includegraphics[width=0.95\columnwidth]{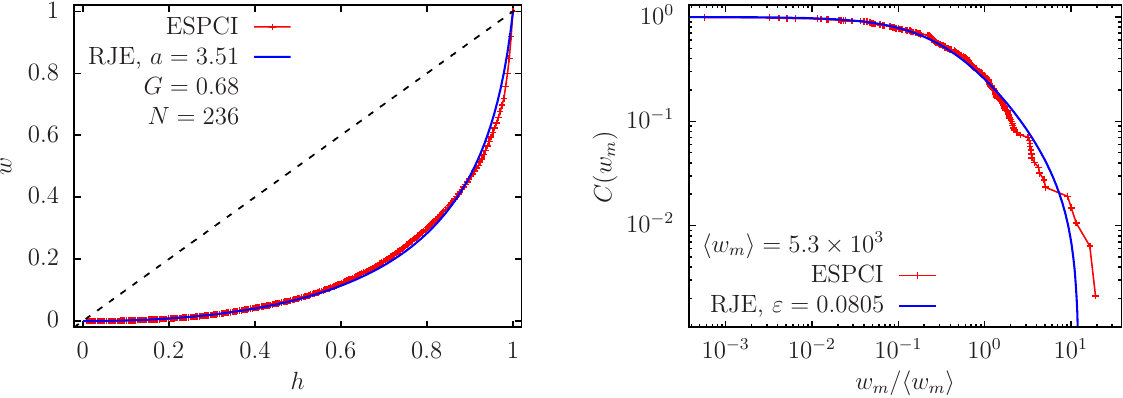}%
\end{center}
\caption{\label{figS21}
\label{fig_ESPCIM500_multi}
As Fig.~\ref{fig1} for ESPCI.
}
\end{figure}

\begin{figure}[h]
\begin{center}
\includegraphics[width=0.95\columnwidth]{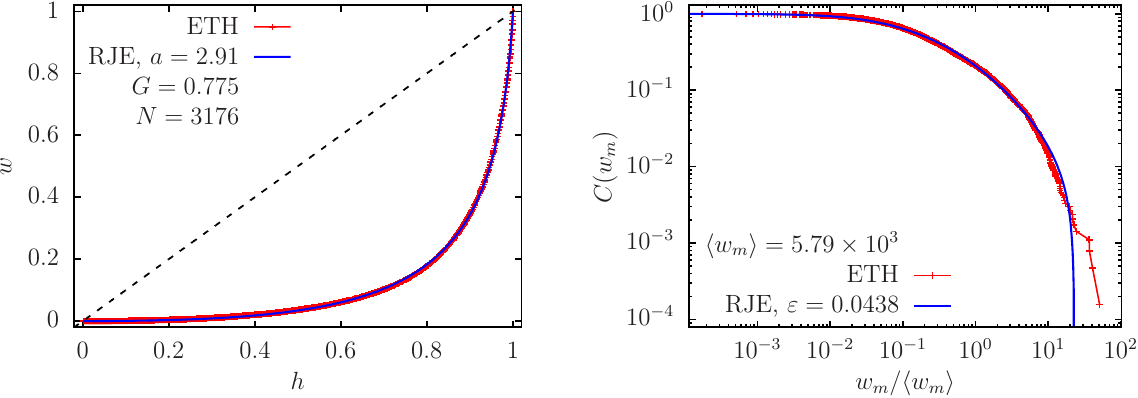}%
\end{center}
\caption{\label{figS22}
\label{fig_ETHM500_multi}
As Fig.~\ref{fig1} for ETH.
}
\end{figure}

\begin{figure}[h]
\begin{center}
\includegraphics[width=0.95\columnwidth]{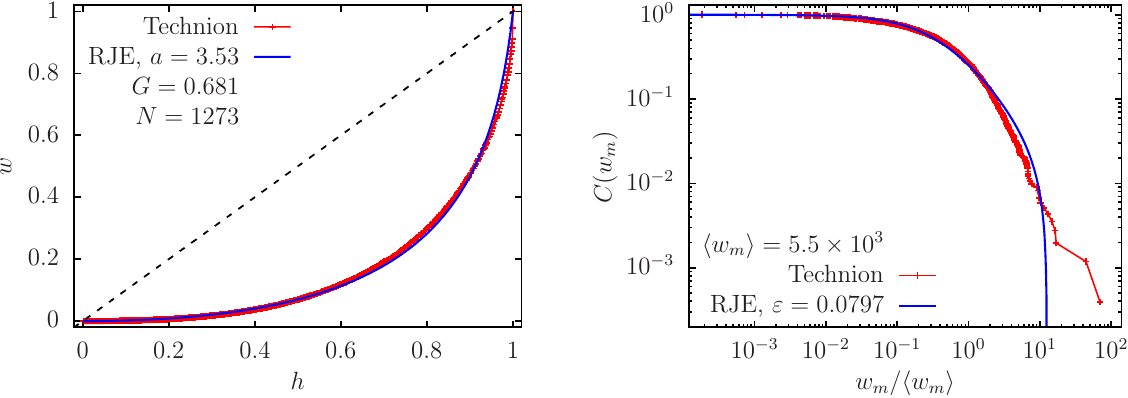}%
\end{center}
\caption{\label{figS23}
\label{fig_TechnionM500_multi}
As Fig.~\ref{fig1} for Technion.
}
\end{figure}

\end{document}